\documentclass[reprint,aps,pre,amsmath,nofootinbib,longbibliography]{revtex4-2}
\usepackage{float}
\usepackage{color}
\usepackage{graphicx}
\usepackage[normalem]{ulem}
\usepackage{epstopdf}
\usepackage{amsfonts}
\usepackage{amsmath}
\usepackage{amssymb}
\usepackage{mathtools}
\usepackage{bbold}
\usepackage{bm}
\usepackage[toc,page]{appendix}
\usepackage{enumerate}
\usepackage{comment}
\newcommand{\x}{\mathbf{x}}
\newcommand{\I}{\mathbb{1}}
\newcommand{\G}{\mathcal{G}}
\newcommand{\A}{\mathbf{A}}
\newcommand{\B}{\mathbf{B}}
\newcommand{\C}{\mathbf{C}}
\newcommand{\Cw}{\mathbf{C}_{\rm w}}
\newcommand{\D}{\mathbf{D}}
\newcommand{\Deff}{\widetilde{\D}}
\newcommand{\eL}{\mathbf{L}}
\newcommand{\J}{\bm{j}}
\newcommand{\V}{\bm{v}}

\newcommand{\AER}{\bm{\mathcal{A}}}
\newcommand{\AERw}{\bm{\mathcal{A}}_{\rm w}}
\newcommand{\Om}{\bm{\Omega}}
\newcommand{\Omw}{\bm{\Omega}_{\rm w}}
\newcommand{\etab}{\bm{\eta}}
\newcommand{\xib}{\bm{\xi}}
\newcommand{\T}{\intercal}

\newcommand\chout{\bgroup\markoverwith{\textcolor{red}{\rule[0.5ex]{2pt}{1.0pt}}}\ULon}
\newcommand\gabout{\bgroup\markoverwith{\textcolor{blue}{\rule[0.5ex]{2pt}{1.0pt}}}\ULon}
\newcommand\grzout{\bgroup\markoverwith{\textcolor{green}{\rule[0.5ex]{2pt}{1.0pt}}}\ULon}

\begin{document}

\title{Irreversibility in linear systems with colored noise}

\author{Grzegorz Gradziuk}
\affiliation{Arnold-Sommerfeld-Center for Theoretical Physics and Center for
  NanoScience, Ludwig-Maximilians-Universit\"at M\"unchen,
   D-80333 M\"unchen, Germany}
   \author{Gabriel Torregrosa}
\affiliation{Arnold-Sommerfeld-Center for Theoretical Physics and Center for
  NanoScience, Ludwig-Maximilians-Universit\"at M\"unchen,
   D-80333 M\"unchen, Germany}
\author{Chase P. Broedersz}
\email{c.p.broedersz@vu.nl}
\affiliation{Arnold-Sommerfeld-Center for Theoretical Physics and Center for
  NanoScience, Ludwig-Maximilians-Universit\"at M\"unchen,
   D-80333 M\"unchen, Germany}
  \affiliation{Department of Physics and Astronomy, Vrije Universiteit Amsterdam, 1081 HV Amsterdam, The Netherlands}  

\date{\today}

\begin{abstract}
Time-irreversibility is a distinctive feature of non-equilibrium dynamics and several measures of irreversibility have been introduced to assess the distance from thermal equilibrium of a stochastically driven system. While the dynamical noise is often approximated as white, in many real applications the time correlations of the random forces can actually be significantly long-lived compared to the relaxation times of the driven system. We analyze the effects of temporal correlations in the noise on commonly used measures of irreversibility and demonstrate how the theoretical framework for white noise driven systems naturally generalizes to the case of colored noise. Specifically, we express the auto-correlation function, the area enclosing rates, and mean phase space velocity in terms of solutions of a Lyapunov equation and in terms of their white noise limit values.
\end{abstract}
\maketitle

\section{Introduction}
\label{sec:intro}
\noindent
The Langevin equation, first introduced to simplify the description of Brownian motion \cite{Langevin1908, Lemons1997}, has become a standard tool for studying stochastic systems in fields ranging from physics, chemistry and electronics \cite{Coffey2017} to climate \cite{Hasselmann1976} and population dynamics \cite{Nisbet1982}. In the original context the random force, or the dynamical noise, was assumed to be delta-correlated in time, owing to the time scale separation between the dynamics of the Brownian particle and the dynamics of fluid molecules.
This assumption was later lifted to account for possible time-correlations in the noise and memory effects, leading to the formulation of a generalized Langevin equation \cite{Mori1965, Kubo1966}. Subsequently, the area of investigation expanded to systems out of thermal equilibrium, both synthetic \cite{Mizuno2007a, Ciliberto2013, Howse2007} and natural ones \cite{Martin2001a, Lau2003, Betz2009, Guo2014, Battle2016, Turlier2016a}, setting up the challenge of quantifying the irreversibility of the dynamics and linking this irreversibility to heat dissipation and entropy production rate \cite{Seifert2012, Roldan2014}. To tackle this problem of measuring the "distance from thermal equilibrium" experimentalists have employed a combination of optical and magnetic tweezers-based microrheology combined with time-lapse microscopy. \cite{Mizuno2007a, Guo2014, Betz2009, Ciliberto2010, Ritort2008, Gnesottoreview}. Simultaneous work on the theoretical side resulted in the foundation of stochastic thermodynamics and discovery of a multitude of fluctuation theorems, exposing the connections between irreversibility and dissipation \cite{Seifert2012, Sekimoto2010}.

Despite the spectacular success of the white noise approximation at modelling the stochastic nature of both equilibrium and non-equilibrium processes, a more complex description is indeed needed when the correlation time of the noise becomes significant, as compared to the natural relaxation times of the analyzed system. In fact, temporal correlations in the noise can qualitatively change the dynamics, e.g., they can lead to stochastic resonance \cite{Gammaitoni1998} and induce phase transitions in the stationary probability distributions \cite{Kitahara1980}.

While increasing the range of applicability and the abundance of phenomena, accounting for time-correlations in the noise largely complicates the mathematical treatment of the dynamics. The loss of Markovianity precludes a straightforward mapping to a Fokker-Planck formalism, in which the time evolution of the probability density and resulting correlations between the degrees of freedom could be calculated. To circumvent this obstacle, numerous approximation schemes and perturbative approaches have been developed for specific regimes of the correlation time of the noise \cite{VanKampen1976, Jung1987, Hanggi1995, Fodor2016a}.
Furthermore, it has been shown, that in certain cases, despite the non-Markovianity of the dynamics, an exact, generalized Fokker-Planck equation with an effective diffusion matrix can be formulated under certain assumptions about the initial preparation of the system, statistics of the noise, or linearity of the dynamics \cite{Hanggi1978, Hanggi1995}.
For a comprehensive review of the topic we refer the reader to \cite{Hanggi1995}.

Here, we abstain from calculating the full probability distribution and instead take a perspective focused on the time-irreversibility of the dynamics. We show how the commonly used measures of irreversibility depend on the correlations of the driving noise, in particular its correlation time. We first introduce a class of linear systems for which analytical solutions can be obtained. Then, we demonstrate how the formulas for the irreversibility measures known for white noise driven systems \cite{Weiss2007, Gonzalez2019, Mura2019a} naturally generalize to the colored noise case and discuss the physical mechanism behind this generalization. Finally, we present an example application of the new formulas to a model system of a driven soft biological network \cite{Mura2019a}, that in the context of irreversibility has up to now only been studied within the white noise approximation.

\begin{figure}
\centering
\includegraphics[width=8cm]{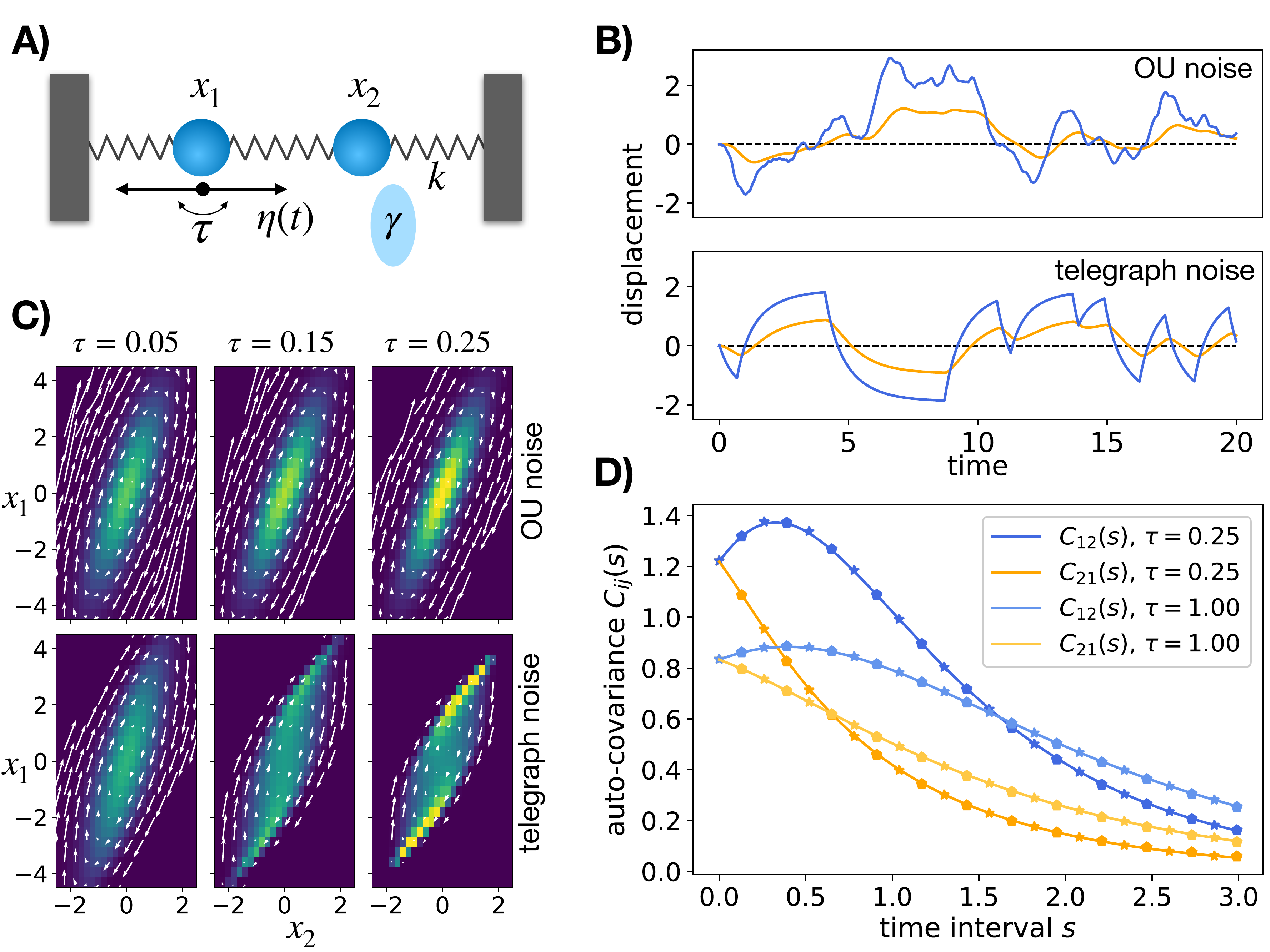}
\caption{
A) Schematics of the two-bead model with bead 1 driven by a colored noise with correlation time $\tau$. B) Sample trajectories of bead displacements $x_1$ (blue) and $x_2$ (orange) for bead 1 driven with telegraph noise (upper plot) and OU noise (lower plot), with $\tau=1$. C) Probability density plots for $(x_1, x_2)$, with OU noise (upper row) and telegraph noise (lower row) for increasing correlation time of the noise. The white arrows represent the mean velocity field. D) Auto-covariance function $C_{ij}(s)=\langle x_i(t)x_j(t+s) \rangle$ for OU noise (stars) and telegraph noise (pentagons).
}
\label{fig:Figura1}
\end{figure}

\section{Linear systems with colored noise}
\label{sec1}
Linear analysis lies at the heart of physics. To reduce the complexity one commonly considers small fluctuations around a fixed point of the deterministic dynamics, for which the restoring force is approximately a linear function of the displacements. In relation to non-equilibrium dynamics and measures of irreversibility, such linear analysis has been applied in diverse contexts: driven biological assemblies \cite{Gladrow2016, Mura2018a, Mura2019a}, population dynamics \cite{McKane2005}, climate oscillations \cite{Hasselmann1976, Weiss2003a, Weiss2020}, or electronic circuits \cite{Ghanta2017, Gonzalez2019}. 

As in these works, we consider an overdamped system for which the deterministic force is a linear function of the instantaneous position in phase space $\x(t)$ at time $t$. The deterministic velocity can thus be written as $\A\x$, where $\A$ includes a constant mobility tensor. The existence of a steady state requires the real parts of all the eigenvalues of $\A$ to be negative. Here, we do not require $\A$ to be symmetric, therefore allowing for non-conservative forces. The Langevin equation then takes the form
\begin{equation}
\dot{\x}(t) = \A\x(t) +\sqrt{2\D_\alpha}\etab_\alpha(t)
\label{Eq:Langevin}
\end{equation}
with possibly colored noise $\etab_\alpha$ characterized by a time-correlation function $\G_\alpha$:
\begin{equation}
\langle \etab_\alpha(t)\rangle=0, 
\quad \langle \etab_\alpha(t)\etab_\beta^\T (t') \rangle =\I \delta_{\alpha,\beta}\G_\alpha(t-t').
\label{Eq:NoiseDef}
\end{equation}
Here, $\alpha$ indexes the pairwise independent sources of the dynamical noise and summation over repeating indices is assumed. The matrix $\D_\alpha$ is an equivalent of the diffusion matrix and describes the amplitude and correlations of the noise acting on different degrees of freedom. These correlations encoded in $\D_\alpha$ can be thought of as spatial correlations of the random forces, if $\x(t)=\{x_i(t)\}$ describes the dynamics of components of a spatially extended system.
We allow in general for a set of pairwise independent noise terms $\{ \sqrt{2\D_\alpha} \etab_\alpha \}$ with corresponding correlations $\{\D_\alpha \}$, $\{\G_\alpha \}$. However, due to the linearity of the system one can always analyse the contributions from each statistically independent noise term separately and then superpose the results. Thereby, for simplicity we consider from now on a single, non-indexed noise term $\sqrt{2\D}\etab(t)$.

It is crucial to note that $\G(t)$ and $\D$ together with Eq.~\eqref{Eq:NoiseDef} do not define the noise uniquely. In fact, two noises characterized by exactly the same $\G(t)$ and $\D$ can lead to qualitatively different dynamics and steady state statistics, as illustrated in Fig.~\ref{fig:Figura1}. On the contrary, the irreversibility measures discussed in this paper, as well as any other two-point correlation function, are indeed fully determined by $\G(t)$ and $\D$, as we shall see later. To illustrate the above points, we present two commonly employed examples of colored noise that share the same time-correlation function.

One of the most common implementations of colored noise defines the dynamics of the noise via an Ornstein-Uhlenbeck (OU) process:
\begin{align}
\dot{\etab}(t)& = -\frac{1}{\tau} \etab(t) + \frac{1}{\tau}\xib(t) \label{Eq:OUnoise} \\
\langle \xi_i(t)\xi_j(t') \rangle &= \delta_{ij}\delta(t-t'),
\end{align}
 where $\xib(t)$ is vector of normalized Gaussian white noises. The time-correlations of $\etab(t)$ decay exponentially:
\begin{equation}
\G(s) = \frac{1}{2\tau} e^{-|s|/\tau},
\label{Eq:ExpCor}
\end{equation}
which allows to identify $\tau$ as the correlation time. A clear advantage of the OU noise is that after including the dynamics of the noise (Eq.~\eqref{Eq:OUnoise}) in the description, one arrives at a Markovian system driven by a Gaussian white noise. This largely simplifies the analysis and for a linear system implies in particular that the conditional probability density $p(\x,t|\x_0,t_0)$ is Gaussian, independently of the value of $\tau$ (see Fig.~\ref{fig:Figura1}). Owing to its simplicity, OU noise is widely employed to model the persistent motion of active Brownian particles \cite{Romanczuk2012, MariniBettoloMarconi2015, Szamel2015, Fodor2016a}.

The assumption about the noise being Gaussian is often legitimate: it is the case, for example, when the random force effectively describes an aggregate action of a large number of factors (such as the collisions of fluid molecules with a Brownian particle), as rationalized by the Central Limit Theorem. 
In other contexts, however, the noise statistics can drastically differ from Gaussian.  As a glaring example, consider a random force that switches between a discrete set of values. Similar stochastic dynamics is exhibited in nature by myosin motors \citep{Mizuno2007a, MacKintosh2008, Brangwynne2008a, Guo2014} exerting step-like contractile forces on the cytoskeleton.
In such systems the noise may be more adequately modelled with a telegraph process \cite{Kac1974}:
\begin{equation}
\eta_i(t) = \frac{(-1)^{n_i(t)}}{\sqrt{2\tau}},
\label{Eq:TeleDef}
\end{equation}
where $\{ n_i(t)\}$ are pairwise independent Poisson processes with transition rate $1/(2\tau)$. The time correlation function $\G(s)$ for the telegraph noise, as defined by Eq.~\eqref{Eq:TeleDef}, is an exponential decay, identical to the one associated to the OU noise (Eq.~\eqref{Eq:ExpCor}). Even though these two types of noise give rise to very different trajectories and steady state probability distributions, they result in identical auto-covariance functions (see Fig.~\ref{fig:Figura1}), which will be discussed in detail in the following section.

Finally, let us remark that we use a convention in which the time correlation function of the noise is normalized to 1, namely: $\int_{-\infty}^{\infty} \G(t)dt = 1$, corresponding to noise of power 1. This choice, implying that $\G(t)\xrightarrow{\tau\to 0} \delta(t)$, allows us to study the effects of transition from white to time-correlated noise, while keeping the power of the noise fixed.

\section{Auto-covariance function}
The auto-covariance function is a standard quantity used in the analysis of time series and in particular the time-irreversibility of the dynamics. Comparing the auto-covariance with the response function can reveal deviations from the Fluctuation-Dissipation theorem \cite{Martin2001, Lau2003, Mizuno2007a, Turlier2016a}, which can be further connected to the heat dissipation rate through the Harada-Sasa relation \cite{Harada2005, Harada2006}. The auto-covariance is defined as
\begin{equation}
\C(t, t+s) = \langle \x(t) \x^\T(t+s) \rangle.
\end{equation}
Under steady state conditions, the covariance function becomes translation invariant in time and we can write $\C(t, t+s)\coloneqq\C(s)$. By definition, the covariance function fulfils $\C^\T(s)=\C(-s)$. For the purpose of calculations, this property allows us to assume without loss of generality that $s>0$, since the covariance for negative time differences $-s$ can be obtained by transposition of $\C(s)$. Time reversibility additionally requires $\C(s) = \C(-s)$, implying that $\C(s)$ must be a symmetric matrix. Note, however, to fulfill $\C(s)=\C^\T(s)$ does not in general require the system to be at equilibrium. In fact, any time correlations of the noise in Eq.~\eqref{Eq:Langevin} not accompanied by a corresponding memory kernel in the deterministic term inevitably violate the Fluctuation-Dissipation Theorem \cite{Kubo1966}.

To derive an equation for $\C(s)$ note that for a linear system one can write a formal solution for the time trajectory:
\begin{equation}
\x(t) = \int_{-\infty}^t e^{\A(t-t')} \sqrt{2\D} \etab(t')dt'
\label{Eq:formalx}
\end{equation}
and for the associated covariance matrix
\begin{equation}
\C(t,t+s)
= \int_{-\infty}^t \!\!\!\! dt' \! \int_{-\infty}^{t+s} \!\!\!\! dt''
e^{\A(t-t')}  2\D\G(t'-t'') e^{\A^\T(t+s-t'')}.
\label{Eq:formalC}
\end{equation}
While mathematically correct, Eq.~\eqref{Eq:formalC} does not provide a simple interpretation of how the time correlations in the noise affect the covariance function.
A more insightful relation is obtained by using the steady state assumption and time translation invariance. Eq.~\eqref{Eq:formalC} combined with the steady state property $\partial_t \C(t, t+s)  = 0$ leads to (see Appendix):
\begin{equation}
\A\C(s) + \C(s)\A^\T = -[\B(s)\D + \D\B^\T(-s)],
\label{Eq:s-Lyapunov_main}
\end{equation}
where $\B(s)$ is defined as
\begin{equation}
\B(s) = 2\int_0^\infty dt e^{\A t}\G(t+s) \quad \forall_{s\in\mathbb{R}} .
\end{equation}
For reasons that will become clear later, we refer to $\B$ as the spreading matrix. First, let us consider a few limiting cases. In the limit of white noise $(\tau\to 0)$ the equal-time covariance matrix $\Cw(s=0)\coloneqq\Cw$ fulfils
\begin{equation}
\A\Cw + \Cw\A^\T = -2\D,
\label{Eq:Lyapunov}
\end{equation}
which is the well known Lyapunov equation \cite{Weiss2003a}. The product $\A\Cw$ can be identified with the matrix of Onsager coefficients $\eL$ and the Lyapunov equation itself with $\eL+\eL^\T=-2\D$ \cite{Onsager1931}.
Comparing Eq.~\eqref{Eq:s-Lyapunov_main} with Eq.~\eqref{Eq:Lyapunov} leads to a key observation: the covariance $\C(s)$ can be calculated by solving a Lyapunov equation with an effective diffusion matrix. This modified diffusion matrix $\Deff(s)$ is set by the spreading matrix $\B(s)$ via
\begin{equation}
\Deff(s) = \frac{1}{2}[\B(s)\D + \D\B^\T(-s)].
\end{equation}
Note that for time irreversible dynamics and $s\neq 0$ the effective diffusion matrix $\Deff(s)$ can in general be non-symmetric, resulting in a covariance function non-symmetric in time. The non-symmetricity of the effective diffusion matrix does not allow for identifying $\Deff(s)$ as a covariance matrix of some effective white noise. Conversely, when considering the equal-time covariance (setting $s=0$), we find a symmetric effective diffusion matrix
\begin{equation}
\Deff = \frac{1}{2}[\B\D + \D\B^\T],
\label{Eq:Deff0}
\end{equation}
which can indeed be interpreted as a covariance matrix of an effective noise. Therefore, the equal-time covariance $\C(s=0)\coloneqq\C$ for a system with time correlated noise and diffusion matrix $\D$ remains unchanged, if one replaces the colored noise with a white noise with an appropriate effective diffusion matrix $\Deff$. In this sense, for linear systems the temporal and spatial correlations in the noise are equivalent at the level of covariance matrix.
This is of particular importance for the case of Gaussian colored noise (e.g. Ornstein-Uhlenbeck noise), for which the stationary probability density is known to be Gaussian. In this case the covariance matrix $\C$ calculated for the system with the effective white noise defines the exact stationary probability distribution of a colored noise system: $p(\x) \sim \exp[-\x^\T\C^{-1}\x/2]$. This special case of Gaussian noise was solved in \cite{Hanggi1995}, where a generalized Fokker-Planck equation for $p(\x, t)$ is derived. In the long time limit the diffusion matrix of the generalized Fokker-Planck equation converges to our effective diffusion matrix $\Deff$ defined as in  Eq.~\eqref{Eq:Deff0}. Importantly, even though with Gaussian noise the stationary probability densities are identical in the coloured- and effective white-noise systems, the actual dynamics can be very different. Moreover, non-Gaussian noise can result in $p(\x)$ qualitatively different from Gaussian (see Fig.~\ref{fig:Figura1}), while, remarkably, the covariance matrix $\C$ is the same as for the effective white-noise system.

\begin{figure}
\centering
\includegraphics[width=8cm]{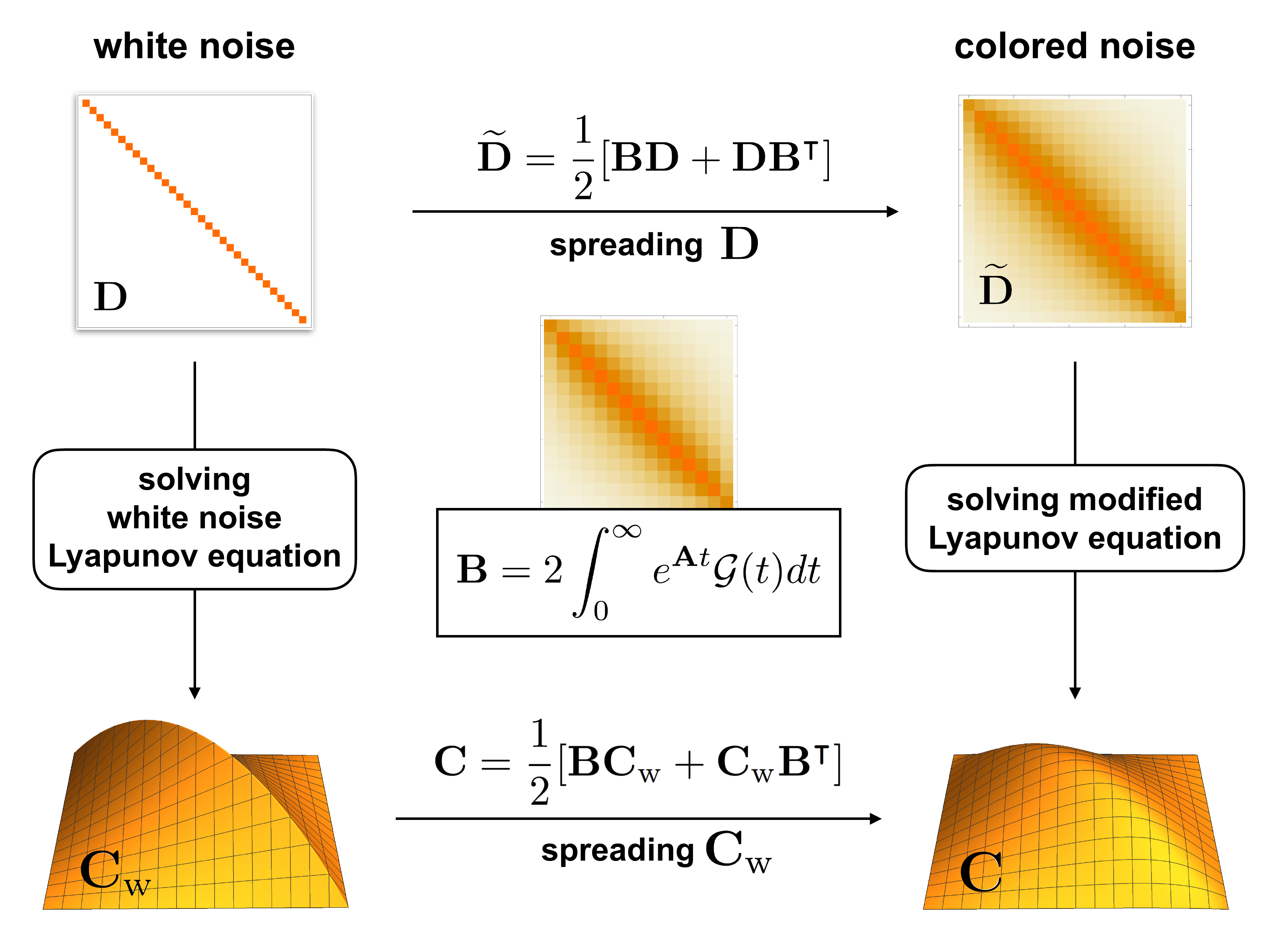}
\caption{
Schematic procedure of obtaining the auto-covariance matrix $\C(s)$ in the colored noise case. Here shown for $s=0$, for a 1-dimensional chain of harmonically coupled beads, with each bead driven by independent noise with correlations decaying exponentially in time. The two possible sequences of steps marked by the arrows lead to equal results for $\C(s)$. These two alternative ways include "spreading" either $\D$, or $\Cw$ using the $\B$ matrix.
}
\label{fig:Figura2}
\end{figure}

Having presented the partial correspondence between temporal and spatial correlations in the noise for linear systems, let us now discuss the physical origin of this connection. Consider two degrees of freedom $x_i$ and $x_j$ that are coupled in some way, directly or indirectly. That is a displacement in $x_i$ results in a displacement of $x_j$, with a magnitude and time dependence encoded by $\A$. When a persistent force is exerted on the $i$th degree of freedom, it is followed by a displacement in $x_j$, which leads to correlations in the force instantaneously experienced by $i$ and $j$. The way these correlations enter in the effective diffusion matrix $\Deff$ is specified by Eq.~\eqref{Eq:Deff0} and the spreading matrix $\B$. The multiplication by $\B$ results in spreading of the elements of $\D$, as exemplified in Fig.~\ref{fig:Figura2}. The directions of spreading are set by the $\A$ matrix, while the range of spreading is controlled by the correlation time $\tau$, or more generally, by the width of $\G(t)$. These two key quantities define $\B$ in terms of an integral, but an explicit expression can be found in certain cases. For the ubiquitous case of exponentially decaying time-correlations, as in Eq.~\eqref{Eq:ExpCor}, one finds
\begin{equation}
\B = (\I -\tau\A)^{-1}.
\end{equation}
The derivation and expressions for $\B(s)$ are presented in the Supplementary Information. Notably, Eq.~\eqref{Eq:s-Lyapunov_main} and the derived expressions for $\B(s)$ are exact and make no assumptions on the noise correlation time $\tau$ being small or large.

We close this section by noting that instead of first calculating the effective diffusion matrix and solving the Lyapunov equation with $\Deff(s)$, one can equivalently first find the equal-time white-noise covariance matrix $\Cw$ (by replacing $\G(s)$ with $\delta(s)$ in the original problem and solving Eq.~\eqref{Eq:Lyapunov}) and calculate the covariance function $\C(s)$ as
\begin{equation}
\C(s) = \frac{1}{2}[ \B(s)\Cw + \Cw\B^\T(-s) ].
\end{equation}
The equivalence of the two approaches is visualised in Fig.~\ref{fig:Figura2} for a specific system comprised of a chain of harmonically coupled overdamped beads, with $s=0$. A proof based on simple algebra and $\A\B=\B\A$ is presented in the appendix. 

\section{Mean velocity and area enclosing rates}
In the previous section we considered two point time covariance functions as indicators of time irreversible dynamics. However, already mean values of instantaneous quantities can reveal time-irreversibility. Consider the instantaneous velocity $\V(t)$ and its mean value at a certain point in phase space $\langle \V(\x) \rangle \coloneqq \langle \V(t) | \x(t)=\x \rangle$. Upon time reversal the velocity changes sign, and therefore time-reversibility requires $\langle \V(\x) \rangle=0$. This fact has been used to define a class of measures of equilibrium based on non-vanishing probability fluxes \cite{Battle2016, Gladrow2016, Mura2018a, Gonzalez2019}. For a linear system driven by white noise the mean phase space velocity is known to be a linear function of $\x$ \cite{Weiss2003a}:
\begin{equation}
\langle\V(\x)\rangle = \Omw \x, \qquad \Omw = (\A+\D\Cw^{-1})
\end{equation}
Note that since for overdamped systems with white noise the velocities are not well defined in terms of the time derivative of the position, one defines $\langle\V(\x)\rangle$ through $\langle\V(\x)\rangle = \J(\x)/p(\x)$, where $\J(\x)$ is the probability current. To avoid the difficulty of measuring the average velocity $\langle \V(\x)\rangle$ at every point in phase space, one can define a more coarse grained measure of irreversibility known as the area enclosing rate \cite{Footnote1}:
\begin{equation}
\AER = \frac{1}{2} \langle \dot{\x} \x^\T - \x \dot{\x}^\T \rangle .
\label{Eq:AERdef}
\end{equation}
This entity can be traced back to the works of Mori and Kubo \cite{Mori1965, Kubo1966}, where it was used in the derivation of the Generalised Langevin Equation. Recently $\AER$ has received renewed attention as a measure of irreversibility \cite{Ghanta2017, Gonzalez2019, Gradziuk2019, Mura2019a,  Weiss2020}, it has been connected to the entropy production rate, and applied in an irreversibility-oriented dimensionality reduction scheme \cite{Gnesotto2020}. Measuring $\AER$ is in fact equivalent to finding a linear expansion of the mean velocity field $\langle\V(\x)\rangle$ \cite{Frishman2020}. It has been demonstrated that for a white noise-driven linear system $\AER$ can be expressed in terms of the covariance matrix as
\begin{equation}
\AERw = \frac{1}{2}[\A\Cw - \Cw\A^\T].
\label{Eq:solAERw}
\end{equation}
Moreover, in this specific case the mean velocity field and area enclosing rates are related via
\begin{equation}
\Omw = \AERw \Cw^{-1}.
\label{Eq:OmrelAER}
\end{equation}
Hitherto, to our knowledge, the computation of $\V(\x)$ and $\AER$ was mostly limited to systems driven by white noise.
Following the same line of thought as for the auto-covariance function, we would like to investigate how the mean phase space velocity and the area enclosing rates change as the correlation time of the noise increases and becomes comparable with the relaxation times of the system, or larger.

Let us begin by considering a linear system driven by an Ornstein-Uhlenbeck noise. In this simple case, the complete dynamics, that include the time evolution of the noise, constitute a white noise linear system. Hence, the probability distribution is Gaussian and consequently the conditional average $\langle\V | \x\rangle$ is also a linear function of $\x$. More specifically, we show that (see Appendix):
\begin{equation}
\langle\V(\x)\rangle = \Om_{\rm OU} \x, \quad \mathrm{with}\quad \Om_{\rm OU} = (\I+\tau\D\Cw^{-1})^{-1}\Omw.
\label{Eq:vOU}
\end{equation}
This expression shows explicitly how the mean velocity $\langle\V(\x)\rangle$ departs from the white noise values as the correlation time of the noise $\tau$ increases. Alternatively, $\Om_{\rm OU}$ can be expressed in a way equivalent to Eq.~\eqref{Eq:OmrelAER}, as demonstrated in the appendix, where we also give an exact expression for the joint probability $p(\x,\V)$.

\begin{figure}
\centering
\includegraphics[width=8cm]{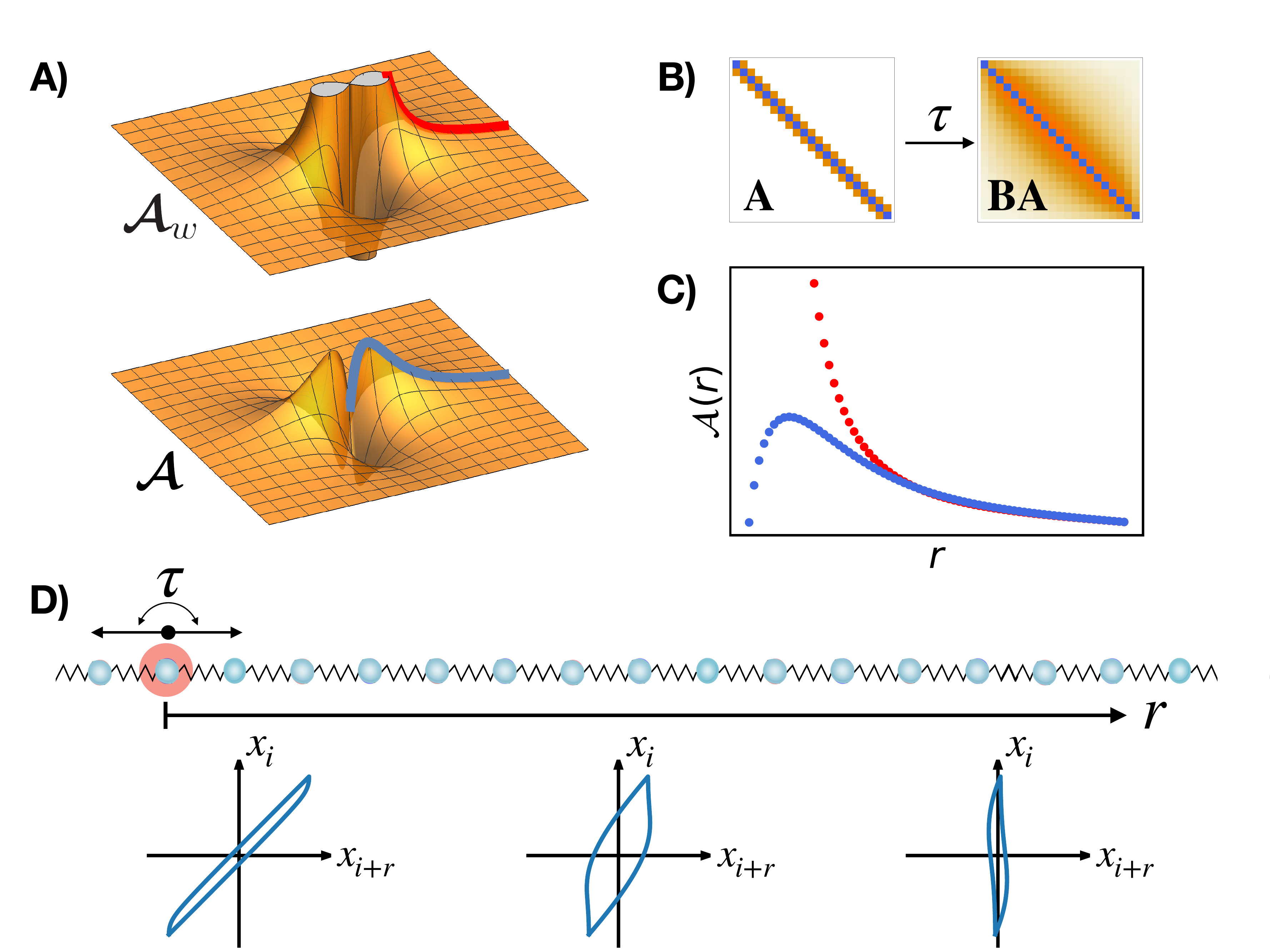}
\caption{
Area enclosing rates for the 1-dimensional chain of harmonically coupled beads, with only the central bead driven. A) Plots of the area enclosing rate matrices for the white noise (top) and the colored noise (bottom) scenarios. The blue and red lines are the same as in panel C). B) The deterministic force matrix $\A$ and the effective force matrix $\A\B$, which enters in Eq.~\eqref{Eq:solAER}. C) Plots of the area enclosing rate for the driven bead and a bead at distance $r$, for white (red) and colored (blue) noise. D) Phase space trajectories of the displacements of bead $i$, driven by a deterministic periodically switching force, and a bead at distance $r$, for increasing values of $r$.
}
\label{fig:Figura3}
\end{figure}

For other types of noise $\langle\V(\x)\rangle$ does not in general show a simple linear dependence on $\x$. Nevertheless, a compact formula can still be found for the area enclosing rates. This is obtained by substituting the formal solution (Eq.~\eqref{Eq:formalx}) into Eq.~\eqref{Eq:AERdef} (for details see the Appendix). One then finds
\begin{equation}
\AER = \frac{1}{2}[\B\A\Cw - \Cw\A^\T\B^\T],
\label{Eq:solAER}
\end{equation}
 that only depends on the correlation function of the noise $\G(s)$ (encoded in $\B$) and on $\D$ (encoded in $\Cw$). This implies, in particular, that two qualitatively different noises (such as OU and telegraph) will yield exactly the same $\AER$, if their two-point correlations given by $\D$ and $\G(s)$ coincide. Eq.~\eqref{Eq:solAER} differs from the white noise formula by $\A$ being replaced with an effective force matrix $\B\A$ (see Fig.~\ref{fig:Figura3}).
 
At this point it is important to emphasise that the results in Eq.~\eqref{Eq:vOU} and Eq.~\eqref{Eq:solAER} are not perturbative and hold for arbitrarily large values of the correlation time $\tau$. Therefore, they allow us to study not only the departure from the white noise behaviour, but the full $\tau$-dependence.
 
Finally, let us demonstrate a relationship between the area enclosing rates $\AER$ and the auto-covariance function $\C(s)$ discussed in the preceding section. As discussed in \cite{Gonzalez2019}, when dealing with a discrete-time signal with time resolution $1/\Delta t$, one has to approximate the velocity in Eq.~\eqref{Eq:AERdef} with a finite difference, leading to:
\begin{align}
\mathcal{A}_{ij} = \frac{1}{2\Delta t}\langle &[x_i(t+\Delta t) - x_i(t) ]x_j(t)  \nonumber \\
 &- x_i(t)[x_j(t+\Delta t) - x_j(t) ]  \rangle  \nonumber \\
  &= \frac{1}{2\Delta t} ( C_{ij}(\Delta t) - C_{ij}(-\Delta t) ).
  \label{Eq:discAER}
\end{align}
For systems driven by time correlated noise only, for which the auto-covariance function is differentiable, we can conclude that $\AER = \frac{d}{ds}\C(s)|_{s=0}$. Consistently, for time reversible dynamics, $\C(s)$ is an even function with derivative 0 at $s=0$. When the dynamics includes a white noise component, $\C(s)$ is no longer differentiable at $s=0$, yet the symmetrized difference in Eq.~\eqref{Eq:discAER} converges to a finite value in the limit of $\Delta t \to 0$.

\section{Example application and emergent length-scale}
In this last section we show how the framework developed in this article can be applied to a physical problem and offer insights inaccessible within a white noise approximation. As a working example we use a toy-model that was recently employed to study the non-equilibrium behaviour across length scales in active biological assemblies \cite{Mura2018a, Gradziuk2019, Mura2019a}. There, the viscoelastic medium mimicking the cytoskeleton was modelled as a lattice of overdamped beads connected by springs and embedded in a viscous fluid. The activity of molecular motors was implemented as a collection of dipole forces delta-correlated in time. Here we revisit this model, lift the assumption about white noise activity and discuss the resulting changes in the non-equilibrium behaviour. For simplicity we focus on a 1-dimensional version of this model - a chain of beads.

\begin{figure}
\centering
\includegraphics[width=8cm]{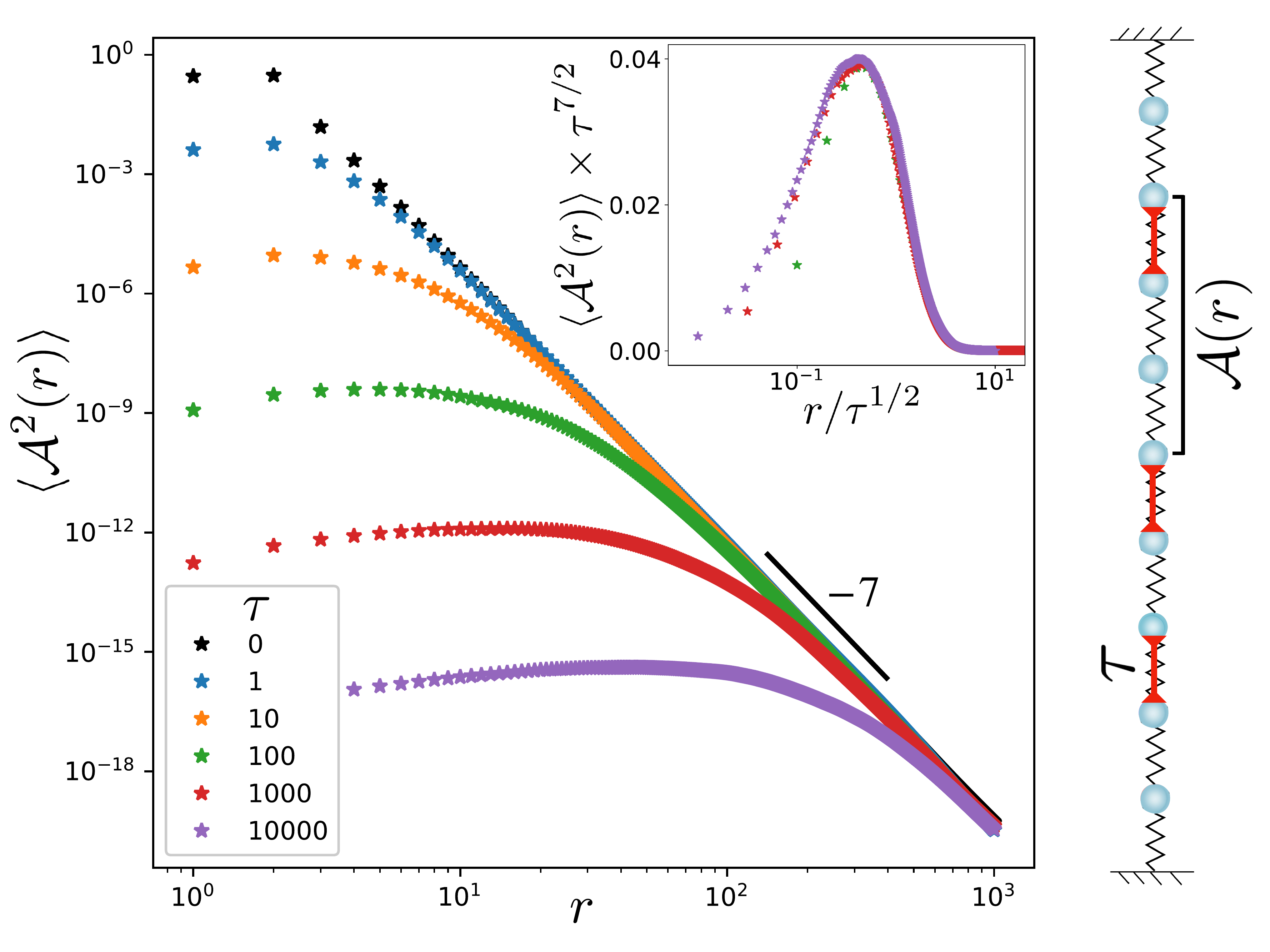}
\caption{
Plots of the mean squared area enclosing rate for pairs of beads at distance $r$. The chain of beads is driven by homogeneously distributed and statistically independent force dipoles with randomly chosen amplitudes. For each dipole the force correlations decay exponentially in time with correlation time $\tau$, as for the telegraph noise. The black markers ($\tau=0$) correspond to white noise. Inset: collapse of the appropriately rescaled curves from the main plot in a linear-log scale. Only the curves for $\tau=\{10^2, 10^3, 10^5\}$, for which the maximum appears at a distance larger than the lattice spacing, are plotted. 
}
\label{fig:Figura4}
\end{figure}

In the previous works it has been shown that with a single activity in the chain, the area enclosing rate measured for two degrees of freedom: the agitated bead $x_i$ and a bead at distance $r$, $x_{i+r}$, displays a power law as a function of the distance $r$. Importantly, with a collection of active agents in the chain, a similar scaling behaviour is observed for the mean squared area enclosing rate $\langle \mathcal{A}^2 (r) \rangle$ measured for pairs of beads at distance $r$ from each other (see Fig.~\ref{fig:Figura4}). Both results assumed the active forces to be delta-correlated in time, or in our formulation $\G(t)=\delta(t)$, and were essentially derived using Eq.~\eqref{Eq:solAERw}.

Now, equipped with the colored noise formula (Eq.~\eqref{Eq:solAER}) we can study the effects of the time correlations of the active forces. We assume force correlations to decay exponentially with time, as in Eq.~\eqref{Eq:ExpCor}, which among others can represent a telegraph process. The results presented in Fig.~\ref{fig:Figura3} (single activity) and in Fig.~\ref{fig:Figura4} (distribution of activities) show a qualitative change in both cases. The functional dependence of $\mathcal{A}_{i, i+r}$ and $\langle \mathcal{A}^2(r) \rangle$ on the distance $r$ is no longer monotonic. At a certain distance the area enclosing rates reach a pronounced maximum followed by a power law decay at large distances. In fact, the behavior for large $r$ coincides with the one predicted for white noise activity - for large distances the relaxation times become much larger than the correlation time of active forces and it is justified to treat the active noise as "white". 
The origin of the maximum can be qualitatively explained in the single activity case with the stereotypical trajectories sketched in Fig.~\ref{fig:Figura3}b). For short distances from the activity both beads experience a large displacement, but since the passive bead reacts almost immediately to the displacement of the driven one, the loops enclosed by the trajectory are very narrow, leading to small $\AER$. At large distances the displacements of the passive bead are small and $\AER$ starts decreasing. A trade-off between the size of displacements and the delay between them gives rise to a maximum at moderate values of $r$. The exact position of the maximum turns out to scale as $\sqrt{\tau}$, as confirmed by the collapse in Fig.~\ref{fig:Figura4}. We suspect this specific scaling, $r_{\rm max}\sim\sqrt{\tau}$, to be related to the distance up to which displacements propagate when applying a point force persistently over time $\tau$. Despite the specificity of the considered system, the simple qualitative explanation of the origin of the maximum suggests that analogous behaviour can be expected in other scenarios. Although it's difficult to argue for a biological relevance of $\AER$ reaching a maximum at a certain distance, the fact itself that a time scale $\tau$ gives rise to a specific length scale is rather remarkable.

\section{Conclusions}
In this paper we showed how temporal correlations in the noise can affect the irreversibility of a systems' linear dynamics. To this end, we considered a class of linear systems described by a Langevin equation and driven by colored noise classified by its autocovariance, which encodes both temporal and spatial correlations. We demonstrated that the formulas commonly employed to calculate standard measures of irreversibility, such as the autocovariance or the area enclosing rates, naturally generalize to the case of colored noise. The expressions we derive are exact and valid for arbitrary correlations of the noise, allowing to study the full dependence of the irreversibility measures on the correlation time of the noise, and not only the departure from the white noise limit.

We discussed how noise scenarios that share the same two-point correlations can lead to drastically different dynamics and steady state probability distribution, while yielding equal results for certain measures of irreversibility. We also showed that for any colored noise model there exists a corresponding white noise model with appropriately modified spatial correlations, which lead to the same covariance matrix.
In practice, when modelling a noisy or driven system, the detailed statistics of the noise at play are unknown and one has to model the noise just based on its primary features. In such cases it is crucial to know how robust the results are to changes in the specific implementation of the noise.
For the case of Ornstein-Uhlenbeck noise, customarily employed to model persistent forces, we find exact expressions for the probability distribution and mean phase space velocity in terms of solutions for the white noise limit.

Finally, we presented an example application of the derived formulas and revisited a model for irreversible dynamics in driven biological network, which up to now was only analyzed within the white noise approximation. We now allowed for time correlations in the active driving, leading to a qualitative change of behaviour of the irreversibility measures at moderate distances. The introduced time correlations in the noise give rise to a specific length scale marked by a maximum of the irreversibility measures. The significance of this and analogous length scales potentially emerging in other non-equilibrium contexts remains an open question.

\begin{acknowledgments}
We thank D. Brückner, F. Gnesotto, F. Mura and P. Ronceray for many stimulating discussions. This work was funded by the Deutsche Forschungsgemeinschaft (DFG, German Research Foundation) under Germany’s Excellence Strategy—EXC-2094 - 390783311, by the DFG grant 418389167 and by the DFG Excellence cluster ORIGINS.
\end{acknowledgments}

\bibliography{BDB_colour}

\begin{thebibliography}{52}%
\makeatletter
\providecommand \@ifxundefined [1]{%
 \@ifx{#1\undefined}
}%
\providecommand \@ifnum [1]{%
 \ifnum #1\expandafter \@firstoftwo
 \else \expandafter \@secondoftwo
 \fi
}%
\providecommand \@ifx [1]{%
 \ifx #1\expandafter \@firstoftwo
 \else \expandafter \@secondoftwo
 \fi
}%
\providecommand \natexlab [1]{#1}%
\providecommand \enquote  [1]{``#1''}%
\providecommand \bibnamefont  [1]{#1}%
\providecommand \bibfnamefont [1]{#1}%
\providecommand \citenamefont [1]{#1}%
\providecommand \href@noop [0]{\@secondoftwo}%
\providecommand \href [0]{\begingroup \@sanitize@url \@href}%
\providecommand \@href[1]{\@@startlink{#1}\@@href}%
\providecommand \@@href[1]{\endgroup#1\@@endlink}%
\providecommand \@sanitize@url [0]{\catcode `\\12\catcode `\$12\catcode
  `\&12\catcode `\#12\catcode `\^12\catcode `\_12\catcode `\%12\relax}%
\providecommand \@@startlink[1]{}%
\providecommand \@@endlink[0]{}%
\providecommand \url  [0]{\begingroup\@sanitize@url \@url }%
\providecommand \@url [1]{\endgroup\@href {#1}{\urlprefix }}%
\providecommand \urlprefix  [0]{URL }%
\providecommand \Eprint [0]{\href }%
\providecommand \doibase [0]{https://doi.org/}%
\providecommand \selectlanguage [0]{\@gobble}%
\providecommand \bibinfo  [0]{\@secondoftwo}%
\providecommand \bibfield  [0]{\@secondoftwo}%
\providecommand \translation [1]{[#1]}%
\providecommand \BibitemOpen [0]{}%
\providecommand \bibitemStop [0]{}%
\providecommand \bibitemNoStop [0]{.\EOS\space}%
\providecommand \EOS [0]{\spacefactor3000\relax}%
\providecommand \BibitemShut  [1]{\csname bibitem#1\endcsname}%
\let\auto@bib@innerbib\@empty
\bibitem [{\citenamefont {Langevin}(1908)}]{Langevin1908}%
  \BibitemOpen
  \bibfield  {author} {\bibinfo {author} {\bibfnamefont {P.}~\bibnamefont
  {Langevin}},\ }\href@noop {} {\bibfield  {journal} {\bibinfo  {journal} {C.
  R. Acad. Sci.}\ }\textbf {\bibinfo {volume} {146}},\ \bibinfo {pages} {530}
  (\bibinfo {year} {1908})}\BibitemShut {NoStop}%
\bibitem [{\citenamefont {Lemons}\ and\ \citenamefont
  {Gythiel}(1997)}]{Lemons1997}%
  \BibitemOpen
  \bibfield  {author} {\bibinfo {author} {\bibfnamefont {D.~S.}\ \bibnamefont
  {Lemons}}\ and\ \bibinfo {author} {\bibfnamefont {A.}~\bibnamefont
  {Gythiel}},\ }\href {https://doi.org/10.1119/1.18725} {\bibfield  {journal}
  {\bibinfo  {journal} {American Journal of Physics}\ }\textbf {\bibinfo
  {volume} {65}},\ \bibinfo {pages} {1079} (\bibinfo {year}
  {1997})}\BibitemShut {NoStop}%
\bibitem [{\citenamefont {Coffey}\ \emph {et~al.}(1996)\citenamefont {Coffey},
  \citenamefont {Kalmykov},\ and\ \citenamefont {Waldron}}]{Coffey2017}%
  \BibitemOpen
  \bibfield  {author} {\bibinfo {author} {\bibfnamefont {W.~T.}\ \bibnamefont
  {Coffey}}, \bibinfo {author} {\bibfnamefont {Y.~P.}\ \bibnamefont
  {Kalmykov}},\ and\ \bibinfo {author} {\bibfnamefont {J.~T.}\ \bibnamefont
  {Waldron}},\ }\href {https://doi.org/10.1142/10490} {\emph {\bibinfo {title}
  {The Langevin Equation}}}\ (\bibinfo  {publisher} {World Scientific,
  Singapore},\ \bibinfo {year} {1996})\BibitemShut {NoStop}%
\bibitem [{\citenamefont {Hasselmann}(1976)}]{Hasselmann1976}%
  \BibitemOpen
  \bibfield  {author} {\bibinfo {author} {\bibfnamefont {K.}~\bibnamefont
  {Hasselmann}},\ }\href {https://doi.org/10.3402/tellusa.v28i6.11316}
  {\bibfield  {journal} {\bibinfo  {journal} {Tellus}\ }\textbf {\bibinfo
  {volume} {28}},\ \bibinfo {pages} {473} (\bibinfo {year} {1976})}\BibitemShut
  {NoStop}%
\bibitem [{\citenamefont {Nisbet}\ and\ \citenamefont
  {Gurney}(1982)}]{Nisbet1982}%
  \BibitemOpen
  \bibfield  {author} {\bibinfo {author} {\bibfnamefont {R.~M.}\ \bibnamefont
  {Nisbet}}\ and\ \bibinfo {author} {\bibfnamefont {W.~S.~C.}\ \bibnamefont
  {Gurney}},\ }\href@noop {} {\emph {\bibinfo {title} {{Modeling Fluctuating
  Populations}}}}\ (\bibinfo  {publisher} {Wiley, New York},\ \bibinfo {year}
  {1982})\BibitemShut {NoStop}%
\bibitem [{\citenamefont {Mori}(1965)}]{Mori1965}%
  \BibitemOpen
  \bibfield  {author} {\bibinfo {author} {\bibfnamefont {H.}~\bibnamefont
  {Mori}},\ }\href@noop {} {\bibfield  {journal} {\bibinfo  {journal} {Progress
  of Theoretical Physics}\ }\textbf {\bibinfo {volume} {33}} (\bibinfo {year}
  {1965})}\BibitemShut {NoStop}%
\bibitem [{\citenamefont {Kubo}(1966)}]{Kubo1966}%
  \BibitemOpen
  \bibfield  {author} {\bibinfo {author} {\bibfnamefont {R.}~\bibnamefont
  {Kubo}},\ }\href {http://iopscience.iop.org/0034-4885/29/1/306} {\bibfield
  {journal} {\bibinfo  {journal} {Rep. Prog. Phys}\ }\textbf {\bibinfo {volume}
  {29}},\ \bibinfo {pages} {255} (\bibinfo {year} {1966})}\BibitemShut
  {NoStop}%
\bibitem [{\citenamefont {Mizuno}\ \emph {et~al.}(2007)\citenamefont {Mizuno},
  \citenamefont {Tardin}, \citenamefont {Schmidt},\ and\ \citenamefont
  {MacKintosh}}]{Mizuno2007a}%
  \BibitemOpen
  \bibfield  {author} {\bibinfo {author} {\bibfnamefont {D.}~\bibnamefont
  {Mizuno}}, \bibinfo {author} {\bibfnamefont {C.}~\bibnamefont {Tardin}},
  \bibinfo {author} {\bibfnamefont {C.~F.}\ \bibnamefont {Schmidt}},\ and\
  \bibinfo {author} {\bibfnamefont {F.~C.}\ \bibnamefont {MacKintosh}},\ }\href
  {https://doi.org/10.1126/science.1134404} {\bibfield  {journal} {\bibinfo
  {journal} {Science}\ }\textbf {\bibinfo {volume} {315}},\ \bibinfo {pages}
  {370} (\bibinfo {year} {2007})}\BibitemShut {NoStop}%
\bibitem [{\citenamefont {Ciliberto}\ \emph {et~al.}(2013)\citenamefont
  {Ciliberto}, \citenamefont {Imparato}, \citenamefont {Naert},\ and\
  \citenamefont {Tanase}}]{Ciliberto2013}%
  \BibitemOpen
  \bibfield  {author} {\bibinfo {author} {\bibfnamefont {S.}~\bibnamefont
  {Ciliberto}}, \bibinfo {author} {\bibfnamefont {A.}~\bibnamefont {Imparato}},
  \bibinfo {author} {\bibfnamefont {A.}~\bibnamefont {Naert}},\ and\ \bibinfo
  {author} {\bibfnamefont {M.}~\bibnamefont {Tanase}},\ }\href
  {https://doi.org/10.1103/PhysRevLett.110.180601} {\bibfield  {journal}
  {\bibinfo  {journal} {Physical Review Letters}\ }\textbf {\bibinfo {volume}
  {110}},\ \bibinfo {pages} {1} (\bibinfo {year} {2013})}\BibitemShut {NoStop}%
\bibitem [{\citenamefont {Howse}\ \emph {et~al.}(2007)\citenamefont {Howse},
  \citenamefont {Jones}, \citenamefont {Ryan}, \citenamefont {Gough},
  \citenamefont {Vafabakhsh},\ and\ \citenamefont {Golestanian}}]{Howse2007}%
  \BibitemOpen
  \bibfield  {author} {\bibinfo {author} {\bibfnamefont {J.~R.}\ \bibnamefont
  {Howse}}, \bibinfo {author} {\bibfnamefont {R.~A.}\ \bibnamefont {Jones}},
  \bibinfo {author} {\bibfnamefont {A.~J.}\ \bibnamefont {Ryan}}, \bibinfo
  {author} {\bibfnamefont {T.}~\bibnamefont {Gough}}, \bibinfo {author}
  {\bibfnamefont {R.}~\bibnamefont {Vafabakhsh}},\ and\ \bibinfo {author}
  {\bibfnamefont {R.}~\bibnamefont {Golestanian}},\ }\href
  {https://doi.org/10.1103/PhysRevLett.99.048102} {\bibfield  {journal}
  {\bibinfo  {journal} {Physical Review Letters}\ }\textbf {\bibinfo {volume}
  {99}},\ \bibinfo {pages} {8} (\bibinfo {year} {2007})},\ \Eprint
  {https://arxiv.org/abs/0706.4406} {arXiv:0706.4406} \BibitemShut {NoStop}%
\bibitem [{\citenamefont {Martin}\ \emph
  {et~al.}(2001{\natexlab{a}})\citenamefont {Martin}, \citenamefont
  {Hudspeth},\ and\ \citenamefont {J{\"{u}}licher}}]{Martin2001a}%
  \BibitemOpen
  \bibfield  {author} {\bibinfo {author} {\bibfnamefont {P.}~\bibnamefont
  {Martin}}, \bibinfo {author} {\bibfnamefont {A.~J.}\ \bibnamefont
  {Hudspeth}},\ and\ \bibinfo {author} {\bibfnamefont {F.}~\bibnamefont
  {J{\"{u}}licher}},\ }\href {https://doi.org/10.1073/pnas.251530598}
  {\bibfield  {journal} {\bibinfo  {journal} {Proceedings of the National
  Academy of Sciences of the United States of America}\ }\textbf {\bibinfo
  {volume} {98}},\ \bibinfo {pages} {14380} (\bibinfo {year}
  {2001}{\natexlab{a}})}\BibitemShut {NoStop}%
\bibitem [{\citenamefont {Lau}\ \emph {et~al.}(2003)\citenamefont {Lau},
  \citenamefont {Hoffman}, \citenamefont {Davies}, \citenamefont {Crocker},\
  and\ \citenamefont {Lubensky}}]{Lau2003}%
  \BibitemOpen
  \bibfield  {author} {\bibinfo {author} {\bibfnamefont {A.~W.~C.}\
  \bibnamefont {Lau}}, \bibinfo {author} {\bibfnamefont {B.~D.}\ \bibnamefont
  {Hoffman}}, \bibinfo {author} {\bibfnamefont {A.}~\bibnamefont {Davies}},
  \bibinfo {author} {\bibfnamefont {J.~C.}\ \bibnamefont {Crocker}},\ and\
  \bibinfo {author} {\bibfnamefont {T.~C.}\ \bibnamefont {Lubensky}},\ }\href
  {https://doi.org/10.1103/PhysRevLett.91.198101} {\bibfield  {journal}
  {\bibinfo  {journal} {Phys. Rev. Lett.}\ }\textbf {\bibinfo {volume} {91}},\
  \bibinfo {pages} {198101} (\bibinfo {year} {2003})}\BibitemShut {NoStop}%
\bibitem [{\citenamefont {Betz}\ \emph {et~al.}(2009)\citenamefont {Betz},
  \citenamefont {Lenz}, \citenamefont {Joanny},\ and\ \citenamefont
  {Sykes}}]{Betz2009}%
  \BibitemOpen
  \bibfield  {author} {\bibinfo {author} {\bibfnamefont {T.}~\bibnamefont
  {Betz}}, \bibinfo {author} {\bibfnamefont {M.}~\bibnamefont {Lenz}}, \bibinfo
  {author} {\bibfnamefont {J.-F.}\ \bibnamefont {Joanny}},\ and\ \bibinfo
  {author} {\bibfnamefont {C.}~\bibnamefont {Sykes}},\ }\href
  {https://doi.org/10.1073/pnas.0904614106} {\bibfield  {journal} {\bibinfo
  {journal} {Proceedings of the National Academy of Sciences}\ }\textbf
  {\bibinfo {volume} {106}},\ \bibinfo {pages} {15320} (\bibinfo {year}
  {2009})}\BibitemShut {NoStop}%
\bibitem [{\citenamefont {Guo}\ \emph {et~al.}(2014)\citenamefont {Guo},
  \citenamefont {Ehrlicher}, \citenamefont {Jensen}, \citenamefont {Renz},
  \citenamefont {Moore}, \citenamefont {Goldman}, \citenamefont
  {Lippincott-Schwartz}, \citenamefont {Mackintosh},\ and\ \citenamefont
  {Weitz}}]{Guo2014}%
  \BibitemOpen
  \bibfield  {author} {\bibinfo {author} {\bibfnamefont {M.}~\bibnamefont
  {Guo}}, \bibinfo {author} {\bibfnamefont {A.~J.}\ \bibnamefont {Ehrlicher}},
  \bibinfo {author} {\bibfnamefont {M.~H.}\ \bibnamefont {Jensen}}, \bibinfo
  {author} {\bibfnamefont {M.}~\bibnamefont {Renz}}, \bibinfo {author}
  {\bibfnamefont {J.~R.}\ \bibnamefont {Moore}}, \bibinfo {author}
  {\bibfnamefont {R.~D.}\ \bibnamefont {Goldman}}, \bibinfo {author}
  {\bibfnamefont {J.}~\bibnamefont {Lippincott-Schwartz}}, \bibinfo {author}
  {\bibfnamefont {F.~C.}\ \bibnamefont {Mackintosh}},\ and\ \bibinfo {author}
  {\bibfnamefont {D.~A.}\ \bibnamefont {Weitz}},\ }\href@noop {} {\bibfield
  {journal} {\bibinfo  {journal} {Cell}\ }\textbf {\bibinfo {volume} {158}},\
  \bibinfo {pages} {822} (\bibinfo {year} {2014})}\BibitemShut {NoStop}%
\bibitem [{\citenamefont {Battle}\ \emph {et~al.}(2016)\citenamefont {Battle},
  \citenamefont {Broedersz}, \citenamefont {Fakhri}, \citenamefont {Geyer},
  \citenamefont {Howard}, \citenamefont {Schmidt},\ and\ \citenamefont
  {MacKintosh}}]{Battle2016}%
  \BibitemOpen
  \bibfield  {author} {\bibinfo {author} {\bibfnamefont {C.}~\bibnamefont
  {Battle}}, \bibinfo {author} {\bibfnamefont {C.~P.}\ \bibnamefont
  {Broedersz}}, \bibinfo {author} {\bibfnamefont {N.}~\bibnamefont {Fakhri}},
  \bibinfo {author} {\bibfnamefont {V.~F.}\ \bibnamefont {Geyer}}, \bibinfo
  {author} {\bibfnamefont {J.}~\bibnamefont {Howard}}, \bibinfo {author}
  {\bibfnamefont {C.~F.}\ \bibnamefont {Schmidt}},\ and\ \bibinfo {author}
  {\bibfnamefont {F.~C.}\ \bibnamefont {MacKintosh}},\ }\href
  {https://doi.org/10.1126/science.aac8167} {\bibfield  {journal} {\bibinfo
  {journal} {Science}\ }\textbf {\bibinfo {volume} {352}},\ \bibinfo {pages}
  {604} (\bibinfo {year} {2016})}\BibitemShut {NoStop}%
\bibitem [{\citenamefont {Turlier}\ \emph {et~al.}(2016)\citenamefont
  {Turlier}, \citenamefont {Fedosov}, \citenamefont {Audoly}, \citenamefont
  {Auth}, \citenamefont {Gov}, \citenamefont {Sykes}, \citenamefont {Joanny},
  \citenamefont {Gompper},\ and\ \citenamefont {Betz}}]{Turlier2016a}%
  \BibitemOpen
  \bibfield  {author} {\bibinfo {author} {\bibfnamefont {H.}~\bibnamefont
  {Turlier}}, \bibinfo {author} {\bibfnamefont {D.~A.}\ \bibnamefont
  {Fedosov}}, \bibinfo {author} {\bibfnamefont {B.}~\bibnamefont {Audoly}},
  \bibinfo {author} {\bibfnamefont {T.}~\bibnamefont {Auth}}, \bibinfo {author}
  {\bibfnamefont {N.~S.}\ \bibnamefont {Gov}}, \bibinfo {author} {\bibfnamefont
  {C.}~\bibnamefont {Sykes}}, \bibinfo {author} {\bibfnamefont {J.~F.}\
  \bibnamefont {Joanny}}, \bibinfo {author} {\bibfnamefont {G.}~\bibnamefont
  {Gompper}},\ and\ \bibinfo {author} {\bibfnamefont {T.}~\bibnamefont
  {Betz}},\ }\href {https://doi.org/10.1038/nphys3621} {\bibfield  {journal}
  {\bibinfo  {journal} {Nature Physics}\ }\textbf {\bibinfo {volume} {12}},\
  \bibinfo {pages} {513} (\bibinfo {year} {2016})}\BibitemShut {NoStop}%
\bibitem [{\citenamefont {Seifert}(2012)}]{Seifert2012}%
  \BibitemOpen
  \bibfield  {author} {\bibinfo {author} {\bibfnamefont {U.}~\bibnamefont
  {Seifert}},\ }\href {https://doi.org/10.1088/0034-4885/75/12/126001}
  {\bibfield  {journal} {\bibinfo  {journal} {Reports on Progress in Physics}\
  }\textbf {\bibinfo {volume} {75}},\ \bibinfo {pages} {126001} (\bibinfo
  {year} {2012})},\ \Eprint {https://arxiv.org/abs/1205.4176} {arXiv:1205.4176}
  \BibitemShut {NoStop}%
\bibitem [{\citenamefont {Rold{\'{a}}n}(2014)}]{Roldan2014}%
  \BibitemOpen
  \bibfield  {author} {\bibinfo {author} {\bibfnamefont {{\'{E}}.}~\bibnamefont
  {Rold{\'{a}}n}},\ }\href
  {https://doi.org/https://doi.org/10.1007/978-3-319-07079-7} {\emph {\bibinfo
  {title} {{Irreversibility and Dissipation in Microscopic Systems}}}}\
  (\bibinfo  {publisher} {Springer, Cham},\ \bibinfo {year} {2014})\BibitemShut
  {NoStop}%
\bibitem [{\citenamefont {Ciliberto}\ \emph {et~al.}(2010)\citenamefont
  {Ciliberto}, \citenamefont {Joubaud},\ and\ \citenamefont
  {Petrosyan}}]{Ciliberto2010}%
  \BibitemOpen
  \bibfield  {author} {\bibinfo {author} {\bibfnamefont {S.}~\bibnamefont
  {Ciliberto}}, \bibinfo {author} {\bibfnamefont {S.}~\bibnamefont {Joubaud}},\
  and\ \bibinfo {author} {\bibfnamefont {A.}~\bibnamefont {Petrosyan}},\
  }\href@noop {} {\bibfield  {journal} {\bibinfo  {journal} {Journal of
  Statistical Mechanics: Theory and Experiment}\ }\textbf {\bibinfo {volume}
  {2010}} (\bibinfo {year} {2010})},\ \Eprint {https://arxiv.org/abs/1009.3362}
  {arXiv:1009.3362} \BibitemShut {NoStop}%
\bibitem [{\citenamefont {Ritort}(2008)}]{Ritort2008}%
  \BibitemOpen
  \bibfield  {author} {\bibinfo {author} {\bibfnamefont {F.}~\bibnamefont
  {Ritort}},\ }\href {https://doi.org/10.1002/9780470238080.ch2} {\bibfield
  {journal} {\bibinfo  {journal} {Advances in Chemical Physics}\ }\textbf
  {\bibinfo {volume} {137}},\ \bibinfo {pages} {31} (\bibinfo {year} {2008})},\
  \Eprint {https://arxiv.org/abs/0705.0455} {arXiv:0705.0455} \BibitemShut
  {NoStop}%
\bibitem [{\citenamefont {Gnesotto}\ \emph {et~al.}(2018)\citenamefont
  {Gnesotto}, \citenamefont {Mura}, \citenamefont {Gladrow},\ and\
  \citenamefont {Broedersz}}]{Gnesottoreview}%
  \BibitemOpen
  \bibfield  {author} {\bibinfo {author} {\bibfnamefont {F.~S.}\ \bibnamefont
  {Gnesotto}}, \bibinfo {author} {\bibfnamefont {F.}~\bibnamefont {Mura}},
  \bibinfo {author} {\bibfnamefont {J.}~\bibnamefont {Gladrow}},\ and\ \bibinfo
  {author} {\bibfnamefont {C.~P.}\ \bibnamefont {Broedersz}},\ }\href
  {https://doi.org/10.1088/1361-6633/aab3ed} {\bibfield  {journal} {\bibinfo
  {journal} {Reports on Progress in Physics}\ }\textbf {\bibinfo {volume}
  {81}},\ \bibinfo {pages} {066601} (\bibinfo {year} {2018})},\ \Eprint
  {https://arxiv.org/abs/1710.03456} {arXiv:1710.03456} \BibitemShut {NoStop}%
\bibitem [{\citenamefont {Sekimoto}(2010)}]{Sekimoto2010}%
  \BibitemOpen
  \bibfield  {author} {\bibinfo {author} {\bibfnamefont {K.}~\bibnamefont
  {Sekimoto}},\ }\href {http://link.springer.com/10.1007/978-3-642-05411-2}
  {\emph {\bibinfo {title} {Stochastic Energetics}}}\ (\bibinfo  {publisher}
  {Springer, Berlin},\ \bibinfo {year} {2010})\BibitemShut {NoStop}%
\bibitem [{\citenamefont {Gammaitoni}\ \emph {et~al.}(1998)\citenamefont
  {Gammaitoni}, \citenamefont {H{\"{a}}nggi}, \citenamefont {Jung},\ and\
  \citenamefont {Marchesoni}}]{Gammaitoni1998}%
  \BibitemOpen
  \bibfield  {author} {\bibinfo {author} {\bibfnamefont {L.}~\bibnamefont
  {Gammaitoni}}, \bibinfo {author} {\bibfnamefont {P.}~\bibnamefont
  {H{\"{a}}nggi}}, \bibinfo {author} {\bibfnamefont {P.}~\bibnamefont {Jung}},\
  and\ \bibinfo {author} {\bibfnamefont {F.}~\bibnamefont {Marchesoni}},\
  }\href {https://doi.org/https://doi.org/10.1103/RevModPhys.70.223} {\bibfield
   {journal} {\bibinfo  {journal} {Reviews of Modern Physics}\ }\textbf
  {\bibinfo {volume} {70}},\ \bibinfo {pages} {223} (\bibinfo {year}
  {1998})}\BibitemShut {NoStop}%
\bibitem [{\citenamefont {Kitahara}\ \emph {et~al.}(1980)\citenamefont
  {Kitahara}, \citenamefont {Horsthemke}, \citenamefont {Lefever},\ and\
  \citenamefont {Inaba}}]{Kitahara1980}%
  \BibitemOpen
  \bibfield  {author} {\bibinfo {author} {\bibfnamefont {K.}~\bibnamefont
  {Kitahara}}, \bibinfo {author} {\bibfnamefont {W.}~\bibnamefont
  {Horsthemke}}, \bibinfo {author} {\bibfnamefont {R.}~\bibnamefont
  {Lefever}},\ and\ \bibinfo {author} {\bibfnamefont {Y.}~\bibnamefont
  {Inaba}},\ }\href {https://doi.org/10.1143/ptp.64.1233} {\bibfield  {journal}
  {\bibinfo  {journal} {Progress of Theoretical Physics}\ }\textbf {\bibinfo
  {volume} {64}},\ \bibinfo {pages} {1233} (\bibinfo {year}
  {1980})}\BibitemShut {NoStop}%
\bibitem [{\citenamefont {{Van Kampen}}(1976)}]{VanKampen1976}%
  \BibitemOpen
  \bibfield  {author} {\bibinfo {author} {\bibfnamefont {N.~G.}\ \bibnamefont
  {{Van Kampen}}},\ }\href {https://doi.org/10.1016/0370-1573(76)90029-6}
  {\bibfield  {journal} {\bibinfo  {journal} {Physics Reports}\ }\textbf
  {\bibinfo {volume} {24}},\ \bibinfo {pages} {171} (\bibinfo {year}
  {1976})}\BibitemShut {NoStop}%
\bibitem [{\citenamefont {Jung}\ and\ \citenamefont
  {H{\"{a}}nggi}(1987)}]{Jung1987}%
  \BibitemOpen
  \bibfield  {author} {\bibinfo {author} {\bibfnamefont {P.}~\bibnamefont
  {Jung}}\ and\ \bibinfo {author} {\bibfnamefont {P.}~\bibnamefont
  {H{\"{a}}nggi}},\ }\href {https://doi.org/10.1103/PhysRevA.35.4464}
  {\bibfield  {journal} {\bibinfo  {journal} {Physical Review A}\ }\textbf
  {\bibinfo {volume} {35}},\ \bibinfo {pages} {4464} (\bibinfo {year}
  {1987})}\BibitemShut {NoStop}%
\bibitem [{\citenamefont {H{\"{a}}nggi}\ and\ \citenamefont
  {Jung}(1995)}]{Hanggi1995}%
  \BibitemOpen
  \bibfield  {author} {\bibinfo {author} {\bibfnamefont {P.}~\bibnamefont
  {H{\"{a}}nggi}}\ and\ \bibinfo {author} {\bibfnamefont {P.}~\bibnamefont
  {Jung}},\ }\href {https://doi.org/10.1002/9780470141489.ch4} {\bibfield
  {journal} {\bibinfo  {journal} {Adv. Chem. Phys.}\ }\textbf {\bibinfo
  {volume} {89}},\ \bibinfo {pages} {239} (\bibinfo {year} {1995})}\BibitemShut
  {NoStop}%
\bibitem [{\citenamefont {Fodor}\ \emph {et~al.}(2016)\citenamefont {Fodor},
  \citenamefont {Nardini}, \citenamefont {Cates}, \citenamefont {Tailleur},
  \citenamefont {Visco},\ and\ \citenamefont {{Van Wijland}}}]{Fodor2016a}%
  \BibitemOpen
  \bibfield  {author} {\bibinfo {author} {\bibfnamefont {{\'{E}}.}~\bibnamefont
  {Fodor}}, \bibinfo {author} {\bibfnamefont {C.}~\bibnamefont {Nardini}},
  \bibinfo {author} {\bibfnamefont {M.~E.}\ \bibnamefont {Cates}}, \bibinfo
  {author} {\bibfnamefont {J.}~\bibnamefont {Tailleur}}, \bibinfo {author}
  {\bibfnamefont {P.}~\bibnamefont {Visco}},\ and\ \bibinfo {author}
  {\bibfnamefont {F.}~\bibnamefont {{Van Wijland}}},\ }\href
  {https://doi.org/10.1103/PhysRevLett.117.038103} {\bibfield  {journal}
  {\bibinfo  {journal} {Physical Review Letters}\ }\textbf {\bibinfo {volume}
  {117}},\ \bibinfo {pages} {1} (\bibinfo {year} {2016})},\ \Eprint
  {https://arxiv.org/abs/1604.00953} {arXiv:1604.00953} \BibitemShut {NoStop}%
\bibitem [{\citenamefont {H{\"{a}}nggi}(1978)}]{Hanggi1978}%
  \BibitemOpen
  \bibfield  {author} {\bibinfo {author} {\bibfnamefont {P.}~\bibnamefont
  {H{\"{a}}nggi}},\ }\href {https://doi.org/10.1007/BF01351552} {\bibfield
  {journal} {\bibinfo  {journal} {Zeitschrift f{\"{u}}r Physik B Condensed
  Matter and Quanta}\ }\textbf {\bibinfo {volume} {31}},\ \bibinfo {pages}
  {407} (\bibinfo {year} {1978})}\BibitemShut {NoStop}%
\bibitem [{\citenamefont {Weiss}(2007)}]{Weiss2007}%
  \BibitemOpen
  \bibfield  {author} {\bibinfo {author} {\bibfnamefont {J.~B.}\ \bibnamefont
  {Weiss}},\ }\href {https://doi.org/10.1103/PhysRevE.76.061128} {\bibfield
  {journal} {\bibinfo  {journal} {Physical Review E - Statistical, Nonlinear,
  and Soft Matter Physics}\ }\textbf {\bibinfo {volume} {76}},\ \bibinfo
  {pages} {1} (\bibinfo {year} {2007})},\ \Eprint
  {https://arxiv.org/abs/0711.2250} {arXiv:0711.2250} \BibitemShut {NoStop}%
\bibitem [{\citenamefont {Gonzalez}\ \emph {et~al.}(2019)\citenamefont
  {Gonzalez}, \citenamefont {Neu},\ and\ \citenamefont
  {Teitsworth}}]{Gonzalez2019}%
  \BibitemOpen
  \bibfield  {author} {\bibinfo {author} {\bibfnamefont {J.~P.}\ \bibnamefont
  {Gonzalez}}, \bibinfo {author} {\bibfnamefont {J.~C.}\ \bibnamefont {Neu}},\
  and\ \bibinfo {author} {\bibfnamefont {S.~W.}\ \bibnamefont {Teitsworth}},\
  }\href {https://doi.org/10.1103/PhysRevE.99.022143} {\bibfield  {journal}
  {\bibinfo  {journal} {Physical Review E}\ }\textbf {\bibinfo {volume} {99}},\
  \bibinfo {pages} {22143} (\bibinfo {year} {2019})}\BibitemShut {NoStop}%
\bibitem [{\citenamefont {Mura}\ \emph {et~al.}(2019)\citenamefont {Mura},
  \citenamefont {Gradziuk},\ and\ \citenamefont {Broedersz}}]{Mura2019a}%
  \BibitemOpen
  \bibfield  {author} {\bibinfo {author} {\bibfnamefont {F.}~\bibnamefont
  {Mura}}, \bibinfo {author} {\bibfnamefont {G.}~\bibnamefont {Gradziuk}},\
  and\ \bibinfo {author} {\bibfnamefont {C.~P.}\ \bibnamefont {Broedersz}},\
  }\href {https://doi.org/10.1039/c9sm01169b} {\bibfield  {journal} {\bibinfo
  {journal} {Soft Matter}\ }\textbf {\bibinfo {volume} {15}},\ \bibinfo {pages}
  {8067} (\bibinfo {year} {2019})},\ \Eprint {https://arxiv.org/abs/1905.13663}
  {arXiv:1905.13663} \BibitemShut {NoStop}%
\bibitem [{\citenamefont {Gladrow}\ \emph {et~al.}(2016)\citenamefont
  {Gladrow}, \citenamefont {Fakhri}, \citenamefont {MacKintosh}, \citenamefont
  {Schmidt},\ and\ \citenamefont {Broedersz}}]{Gladrow2016}%
  \BibitemOpen
  \bibfield  {author} {\bibinfo {author} {\bibfnamefont {J.}~\bibnamefont
  {Gladrow}}, \bibinfo {author} {\bibfnamefont {N.}~\bibnamefont {Fakhri}},
  \bibinfo {author} {\bibfnamefont {F.~C.}\ \bibnamefont {MacKintosh}},
  \bibinfo {author} {\bibfnamefont {C.~F.}\ \bibnamefont {Schmidt}},\ and\
  \bibinfo {author} {\bibfnamefont {C.~P.}\ \bibnamefont {Broedersz}},\ }\href
  {https://doi.org/10.1103/PhysRevLett.116.248301} {\bibfield  {journal}
  {\bibinfo  {journal} {Phys. Rev. Lett.}\ }\textbf {\bibinfo {volume} {116}},\
  \bibinfo {pages} {248301} (\bibinfo {year} {2016})}\BibitemShut {NoStop}%
\bibitem [{\citenamefont {Mura}\ \emph {et~al.}(2018)\citenamefont {Mura},
  \citenamefont {Gradziuk},\ and\ \citenamefont {Broedersz}}]{Mura2018a}%
  \BibitemOpen
  \bibfield  {author} {\bibinfo {author} {\bibfnamefont {F.}~\bibnamefont
  {Mura}}, \bibinfo {author} {\bibfnamefont {G.}~\bibnamefont {Gradziuk}},\
  and\ \bibinfo {author} {\bibfnamefont {C.~P.}\ \bibnamefont {Broedersz}},\
  }\href {https://doi.org/10.1103/PhysRevLett.121.038002} {\bibfield  {journal}
  {\bibinfo  {journal} {Physical Review Letters}\ }\textbf {\bibinfo {volume}
  {121}},\ \bibinfo {pages} {38002} (\bibinfo {year} {2018})},\ \Eprint
  {https://arxiv.org/abs/1803.02797} {arXiv:1803.02797} \BibitemShut {NoStop}%
\bibitem [{\citenamefont {McKane}\ and\ \citenamefont
  {Newman}(2005)}]{McKane2005}%
  \BibitemOpen
  \bibfield  {author} {\bibinfo {author} {\bibfnamefont {A.~J.}\ \bibnamefont
  {McKane}}\ and\ \bibinfo {author} {\bibfnamefont {T.~J.}\ \bibnamefont
  {Newman}},\ }\href {https://doi.org/10.1103/PhysRevLett.94.218102} {\bibfield
   {journal} {\bibinfo  {journal} {Physical Review Letters}\ }\textbf {\bibinfo
  {volume} {94}},\ \bibinfo {pages} {1} (\bibinfo {year} {2005})},\ \Eprint
  {https://arxiv.org/abs/0501023} {arXiv:0501023 [q-bio]} \BibitemShut
  {NoStop}%
\bibitem [{\citenamefont {Weiss}(2003)}]{Weiss2003a}%
  \BibitemOpen
  \bibfield  {author} {\bibinfo {author} {\bibfnamefont {J.~B.}\ \bibnamefont
  {Weiss}},\ }\href {https://doi.org/10.3402/tellusa.v55i3.12093} {\bibfield
  {journal} {\bibinfo  {journal} {Tellus A: Dynamic Meteorology and
  Oceanography}\ }\textbf {\bibinfo {volume} {55}},\ \bibinfo {pages} {208}
  (\bibinfo {year} {2003})}\BibitemShut {NoStop}%
\bibitem [{\citenamefont {Weiss}\ \emph {et~al.}(2020)\citenamefont {Weiss},
  \citenamefont {Fox-Kemper}, \citenamefont {Mandal}, \citenamefont {Arin},\
  and\ \citenamefont {Zia}}]{Weiss2020}%
  \BibitemOpen
  \bibfield  {author} {\bibinfo {author} {\bibfnamefont {J.~B.}\ \bibnamefont
  {Weiss}}, \bibinfo {author} {\bibfnamefont {B.}~\bibnamefont {Fox-Kemper}},
  \bibinfo {author} {\bibfnamefont {D.}~\bibnamefont {Mandal}}, \bibinfo
  {author} {\bibfnamefont {N.~D.}\ \bibnamefont {Arin}},\ and\ \bibinfo
  {author} {\bibfnamefont {R.~K.~P.}\ \bibnamefont {Zia}},\ }\href
  {https://doi.org/10.1007/s10955-019-02394-1} {\bibfield  {journal} {\bibinfo
  {journal} {Journal of Statistical Physics}\ }\textbf {\bibinfo {volume}
  {179}},\ \bibinfo {pages} {1010} (\bibinfo {year} {2020})}\BibitemShut
  {NoStop}%
\bibitem [{\citenamefont {Ghanta}\ \emph {et~al.}(2017)\citenamefont {Ghanta},
  \citenamefont {Neu},\ and\ \citenamefont {Teitsworth}}]{Ghanta2017}%
  \BibitemOpen
  \bibfield  {author} {\bibinfo {author} {\bibfnamefont {A.}~\bibnamefont
  {Ghanta}}, \bibinfo {author} {\bibfnamefont {J.~C.}\ \bibnamefont {Neu}},\
  and\ \bibinfo {author} {\bibfnamefont {S.}~\bibnamefont {Teitsworth}},\
  }\href {https://doi.org/10.1103/PhysRevE.95.032128} {\bibfield  {journal}
  {\bibinfo  {journal} {Physical Review E}\ }\textbf {\bibinfo {volume} {95}},\
  \bibinfo {pages} {1} (\bibinfo {year} {2017})}\BibitemShut {NoStop}%
\bibitem [{\citenamefont {Romanczuk}\ \emph {et~al.}(2012)\citenamefont
  {Romanczuk}, \citenamefont {B{\"{a}}r}, \citenamefont {Ebeling},
  \citenamefont {Lindner},\ and\ \citenamefont
  {Schimansky-Geier}}]{Romanczuk2012}%
  \BibitemOpen
  \bibfield  {author} {\bibinfo {author} {\bibfnamefont {P.}~\bibnamefont
  {Romanczuk}}, \bibinfo {author} {\bibfnamefont {M.}~\bibnamefont
  {B{\"{a}}r}}, \bibinfo {author} {\bibfnamefont {W.}~\bibnamefont {Ebeling}},
  \bibinfo {author} {\bibfnamefont {B.}~\bibnamefont {Lindner}},\ and\ \bibinfo
  {author} {\bibfnamefont {L.}~\bibnamefont {Schimansky-Geier}},\ }\href
  {https://doi.org/10.1140/epjst/e2012-01529-y} {\bibfield  {journal} {\bibinfo
   {journal} {European Physical Journal: Special Topics}\ }\textbf {\bibinfo
  {volume} {202}},\ \bibinfo {pages} {1} (\bibinfo {year} {2012})}\BibitemShut
  {NoStop}%
\bibitem [{\citenamefont {{Marini Bettolo Marconi}}\ and\ \citenamefont
  {Maggi}(2015)}]{MariniBettoloMarconi2015}%
  \BibitemOpen
  \bibfield  {author} {\bibinfo {author} {\bibfnamefont {U.}~\bibnamefont
  {{Marini Bettolo Marconi}}}\ and\ \bibinfo {author} {\bibfnamefont
  {C.}~\bibnamefont {Maggi}},\ }\href {https://doi.org/10.1039/c5sm01718a}
  {\bibfield  {journal} {\bibinfo  {journal} {Soft Matter}\ }\textbf {\bibinfo
  {volume} {11}},\ \bibinfo {pages} {8768} (\bibinfo {year} {2015})},\ \Eprint
  {https://arxiv.org/abs/1507.03443} {arXiv:1507.03443} \BibitemShut {NoStop}%
\bibitem [{\citenamefont {Szamel}\ \emph {et~al.}(2015)\citenamefont {Szamel},
  \citenamefont {Flenner},\ and\ \citenamefont {Berthier}}]{Szamel2015}%
  \BibitemOpen
  \bibfield  {author} {\bibinfo {author} {\bibfnamefont {G.}~\bibnamefont
  {Szamel}}, \bibinfo {author} {\bibfnamefont {E.}~\bibnamefont {Flenner}},\
  and\ \bibinfo {author} {\bibfnamefont {L.}~\bibnamefont {Berthier}},\ }\href
  {https://doi.org/10.1103/PhysRevE.91.062304} {\bibfield  {journal} {\bibinfo
  {journal} {Physical Review E - Statistical, Nonlinear, and Soft Matter
  Physics}\ }\textbf {\bibinfo {volume} {91}},\ \bibinfo {pages} {1} (\bibinfo
  {year} {2015})},\ \Eprint {https://arxiv.org/abs/1501.01333}
  {arXiv:1501.01333} \BibitemShut {NoStop}%
\bibitem [{\citenamefont {MacKintosh}\ and\ \citenamefont
  {Levine}(2008)}]{MacKintosh2008}%
  \BibitemOpen
  \bibfield  {author} {\bibinfo {author} {\bibfnamefont {F.~C.}\ \bibnamefont
  {MacKintosh}}\ and\ \bibinfo {author} {\bibfnamefont {A.~J.}\ \bibnamefont
  {Levine}},\ }\href {https://doi.org/10.1103/PhysRevLett.100.018104}
  {\bibfield  {journal} {\bibinfo  {journal} {Phys. Rev. Lett.}\ }\textbf
  {\bibinfo {volume} {100}},\ \bibinfo {pages} {18104} (\bibinfo {year}
  {2008})},\ \Eprint {https://arxiv.org/abs/0704.3794} {arXiv:0704.3794}
  \BibitemShut {NoStop}%
\bibitem [{\citenamefont {Brangwynne}\ \emph {et~al.}(2008)\citenamefont
  {Brangwynne}, \citenamefont {Koenderink}, \citenamefont {Mackintosh},\ and\
  \citenamefont {Weitz}}]{Brangwynne2008a}%
  \BibitemOpen
  \bibfield  {author} {\bibinfo {author} {\bibfnamefont {C.~P.}\ \bibnamefont
  {Brangwynne}}, \bibinfo {author} {\bibfnamefont {G.~H.}\ \bibnamefont
  {Koenderink}}, \bibinfo {author} {\bibfnamefont {F.~C.}\ \bibnamefont
  {Mackintosh}},\ and\ \bibinfo {author} {\bibfnamefont {D.~A.}\ \bibnamefont
  {Weitz}},\ }\href {https://doi.org/10.1103/PhysRevLett.100.118104} {\bibfield
   {journal} {\bibinfo  {journal} {Phys. Rev. Lett.}\ }\textbf {\bibinfo
  {volume} {100}},\ \bibinfo {pages} {118104} (\bibinfo {year}
  {2008})}\BibitemShut {NoStop}%
\bibitem [{\citenamefont {Kac}(1974)}]{Kac1974}%
  \BibitemOpen
  \bibfield  {author} {\bibinfo {author} {\bibfnamefont {M.}~\bibnamefont
  {Kac}},\ }\href@noop {} {\bibfield  {journal} {\bibinfo  {journal} {The Rocky
  Mountain Journal of Mathematics}\ }\textbf {\bibinfo {volume} {4}},\ \bibinfo
  {pages} {497} (\bibinfo {year} {1974})}\BibitemShut {NoStop}%
\bibitem [{\citenamefont {Martin}\ \emph
  {et~al.}(2001{\natexlab{b}})\citenamefont {Martin}, \citenamefont
  {Hudspeth},\ and\ \citenamefont {J{\"{u}}licher}}]{Martin2001}%
  \BibitemOpen
  \bibfield  {author} {\bibinfo {author} {\bibfnamefont {P.}~\bibnamefont
  {Martin}}, \bibinfo {author} {\bibfnamefont {A.~J.}\ \bibnamefont
  {Hudspeth}},\ and\ \bibinfo {author} {\bibfnamefont {F.}~\bibnamefont
  {J{\"{u}}licher}},\ }\href {https://doi.org/10.1073/pnas.251530598}
  {\bibfield  {journal} {\bibinfo  {journal} {Proceedings of the National
  Academy of Sciences}\ }\textbf {\bibinfo {volume} {98}},\ \bibinfo {pages}
  {14380} (\bibinfo {year} {2001}{\natexlab{b}})}\BibitemShut {NoStop}%
\bibitem [{\citenamefont {Harada}\ and\ \citenamefont
  {Sasa}(2005)}]{Harada2005}%
  \BibitemOpen
  \bibfield  {author} {\bibinfo {author} {\bibfnamefont {T.}~\bibnamefont
  {Harada}}\ and\ \bibinfo {author} {\bibfnamefont {S.~I.}\ \bibnamefont
  {Sasa}},\ }\href {https://doi.org/10.1103/PhysRevLett.95.130602} {\bibfield
  {journal} {\bibinfo  {journal} {Physical Review Letters}\ }\textbf {\bibinfo
  {volume} {95}},\ \bibinfo {pages} {1} (\bibinfo {year} {2005})},\ \Eprint
  {https://arxiv.org/abs/0502505} {arXiv:0502505 [cond-mat]} \BibitemShut
  {NoStop}%
\bibitem [{\citenamefont {Harada}\ and\ \citenamefont
  {Sasa}(2006)}]{Harada2006}%
  \BibitemOpen
  \bibfield  {author} {\bibinfo {author} {\bibfnamefont {T.}~\bibnamefont
  {Harada}}\ and\ \bibinfo {author} {\bibfnamefont {S.~I.}\ \bibnamefont
  {Sasa}},\ }\bibfield  {journal} {\bibinfo  {journal} {Physical Review E -
  Statistical, Nonlinear, and Soft Matter Physics}\ }\textbf {\bibinfo {volume}
  {73}},\ \href {https://doi.org/10.1103/PhysRevE.73.026131}
  {10.1103/PhysRevE.73.026131} (\bibinfo {year} {2006}),\ \Eprint
  {https://arxiv.org/abs/0510723} {arXiv:0510723 [cond-mat]} \BibitemShut
  {NoStop}%
\bibitem [{\citenamefont {Onsager}(1931)}]{Onsager1931}%
  \BibitemOpen
  \bibfield  {author} {\bibinfo {author} {\bibfnamefont {L.}~\bibnamefont
  {Onsager}},\ }\href {https://doi.org/10.3109/09546639309082148} {\bibfield
  {journal} {\bibinfo  {journal} {Physical Review}\ }\textbf {\bibinfo {volume}
  {38}},\ \bibinfo {pages} {2265} (\bibinfo {year} {1931})}\BibitemShut
  {NoStop}%
\bibitem [{Foo()}]{Footnote1}%
  \BibitemOpen
  \href@noop {} {\bibinfo {title} {{This antisymmetrized form is only of
  relevance for the white noise case, where it makes the expression independent
  of the applied integration convention. With colored noise one can
  unambiguously write $\AER=\langle\dot{\x}\x^\T\rangle$.}}}\BibitemShut
  {Stop}%
\bibitem [{\citenamefont {Gradziuk}\ \emph {et~al.}(2019)\citenamefont
  {Gradziuk}, \citenamefont {Mura},\ and\ \citenamefont
  {Broedersz}}]{Gradziuk2019}%
  \BibitemOpen
  \bibfield  {author} {\bibinfo {author} {\bibfnamefont {G.}~\bibnamefont
  {Gradziuk}}, \bibinfo {author} {\bibfnamefont {F.}~\bibnamefont {Mura}},\
  and\ \bibinfo {author} {\bibfnamefont {C.~P.}\ \bibnamefont {Broedersz}},\
  }\href {https://doi.org/10.1103/PhysRevE.99.052406} {\bibfield  {journal}
  {\bibinfo  {journal} {Physical Review E}\ }\textbf {\bibinfo {volume} {99}},\
  \bibinfo {pages} {1} (\bibinfo {year} {2019})},\ \Eprint
  {https://arxiv.org/abs/1901.03132} {arXiv:1901.03132} \BibitemShut {NoStop}%
\bibitem [{\citenamefont {Gnesotto}\ \emph {et~al.}(2020)\citenamefont
  {Gnesotto}, \citenamefont {Gradziuk}, \citenamefont {Ronceray},\ and\
  \citenamefont {Broedersz}}]{Gnesotto2020}%
  \BibitemOpen
  \bibfield  {author} {\bibinfo {author} {\bibfnamefont {F.~S.}\ \bibnamefont
  {Gnesotto}}, \bibinfo {author} {\bibfnamefont {G.}~\bibnamefont {Gradziuk}},
  \bibinfo {author} {\bibfnamefont {P.}~\bibnamefont {Ronceray}},\ and\
  \bibinfo {author} {\bibfnamefont {C.~P.}\ \bibnamefont {Broedersz}},\ }\href
  {https://doi.org/10.1038/s41467-020-18796-9} {\bibfield  {journal} {\bibinfo
  {journal} {Nature Communications}\ }\textbf {\bibinfo {volume} {11}},\
  \bibinfo {pages} {1} (\bibinfo {year} {2020})},\ \Eprint
  {https://arxiv.org/abs/2001.08642} {arXiv:2001.08642} \BibitemShut {NoStop}%
\bibitem [{\citenamefont {Frishman}\ and\ \citenamefont
  {Ronceray}(2020)}]{Frishman2020}%
  \BibitemOpen
  \bibfield  {author} {\bibinfo {author} {\bibfnamefont {A.}~\bibnamefont
  {Frishman}}\ and\ \bibinfo {author} {\bibfnamefont {P.}~\bibnamefont
  {Ronceray}},\ }\href {https://doi.org/10.1103/PhysRevX.10.021009} {\bibfield
  {journal} {\bibinfo  {journal} {Physical Review X}\ }\textbf {\bibinfo
  {volume} {10}},\ \bibinfo {pages} {21009} (\bibinfo {year} {2020})},\ \Eprint
  {https://arxiv.org/abs/1809.09650} {arXiv:1809.09650} \BibitemShut {NoStop}%
\end{thebibliography}%

\onecolumngrid
\begin{appendices}
\section{Derivation of auto-covariance function}
\label{appendix1}
Consider a linear system with a time-correlated driving force described with the following equation of motion:
\begin{equation}
\dot{\x}(t) = \A\x(t) +\sqrt{2\D} \etab(t), \quad 
\text{with} \quad \langle \etab(t)\rangle=0, 
\quad \langle \etab(t)\etab^\T (t') \rangle =\I \G(t-t')
\end{equation}
The formal solution can be written as
\begin{equation}
\x(t) = \int_{-\infty}^t e^{\A(t-t')} \sqrt{2\D} \etab(t')dt'
\end{equation}
Similarly, we can formally express the auto-covariance function as
\begin{equation}
\C(t,t+s) = \langle \x(t)\x^\T (t+s) \rangle 
= \int_{-\infty}^t dt' \int_{-\infty}^{t+s} dt''
e^{\A(t-t')}  2\D\G(t'-t'') e^{\A^\T(t+s-t'')}
\end{equation}
Note that by the definition, at the steady state the auto-covariance function $\C(s)\coloneqq \C(t,t+s)$ has the property $\C(-s)=\C^\T(s)$, so in the following we can assume without loss of generality that $s>0$, unless specified otherwise.
Using the following lemma for differentiation:
\begin{equation}
\partial_t \int_{-\infty}^t dt' \int_{-\infty}^{t+s} dt'' f(t,t',t'')
= \int_{-\infty}^t dt' \int_{-\infty}^{t+s} dt'' \partial_t f(t,t',t'')
+\int_{-\infty}^{t+s} dt'' f(t,t,t'')
+ \int_{-\infty}^t dt' f(t,t',t+s)
\end{equation}
and  using the fact that at the stationary state $\partial_t \C(s) =0$, we derive an equation for the auto-covariance:
\begin{align}
\partial_t \C(t,t+s) = 0 = \A\C(s)+\C(s)\A^\T 
+\int_{-\infty}^{t+s}dt'' 2\D e^{\A^\T (t+s-t'')}\G(t-t'')
+\int_{-\infty}^{t}dt' e^{\A(t-t')}2\D\G(t'-t-s) \\
0=\A\C(s)+\C(s)\A^\T
+\D\underbrace{\int_{-\infty}^{t}dt' 2 e^{\A^\T (t-t')}\G(t-t'-s)}_{\B^\T(-s)}
+\underbrace{\int_{-\infty}^{t}dt' 2e^{\A(t-t')}\G(t-t'+s)}_{\B(s)}  \D  \\
\A\C(s)+\C(s)\A^\T = - [\D\B^\T(-s) + \B(s)\D ] \label{Eq:s-Lyapunov}
\end{align}
 Where we have used the time symmetry of the time correlation function of the noise $\G(t)=\G(-t)$ and defined the 'spreading matrix' $\B(s)$:
\begin{equation}
 \B(s)=\int_0^{\infty}dt' 2e^{\A t'}\G(t'+s) \quad\text{ for all } s\in\mathbb{R}
\end{equation}
It is instructive to consider a set of limiting cases for Eq.~\eqref{Eq:s-Lyapunov}. \\

1. Calculating the covariance matrix $\C\coloneqq \C(0)$ for time-correlated noise:
\begin{equation}
\A\C+\C\A^\T = - (\D\B^\T + \B\D ) \quad \text{with} \quad \B\coloneqq\B(0) 
= 2\int_0^{\infty}dt' e^{\A t'}\G(t').
\label{Eq:Lyap1}
\end{equation}

2. Calculating the auto-covariance function $\Cw(s)$ for white noise ($\G(t)=\delta(t)$):
\begin{equation}
\A\Cw(s)+\Cw(s)\A^\T = - 2\D e^{\A^\T s}
\label{Eq:Lyap2}
\end{equation}

3. Calculating the covariance matrix $\Cw$ for white noise:
\begin{equation}
\A\Cw+\Cw\A^\T = - 2\D
\label{Eq:Lyap3}
\end{equation}
which reduces to solving the standard Lyapunov equation.\\

There is an interesting relation between case 3 and Eq.~\eqref{Eq:s-Lyapunov}. Substituting the expression for $\D$ from Eq.~\eqref{Eq:Lyap3} into Eq.~\eqref{Eq:s-Lyapunov} one obtains
\begin{align}
\A\C(s)+\C(s)\A^\T 
&= \frac{1}{2}\left[ \B(s)\A\Cw + \B(s)\Cw\A^\T + \A\Cw\B^\T(-s) + \Cw\A^\T\B^\T(-s) \right] \\
&= \A \left[ \frac{1}{2}( \B(s)\Cw + \Cw\B^\T(-s) ) \right] + \left[ \frac{1}{2}( \B(s)\Cw + \Cw\B^\T(-s) ) \right] \A^\T
\end{align}
and using the uniqueness of the solution of the Lyapunov equation we conclude that
\begin{equation}
\C(s) = \frac{1}{2}[ \B(s)\Cw + \Cw\B^\T(-s) ] 
\label{Eq:suppC}
\end{equation}

For an exponentially correlated noise, with $\G(t) = \frac{1}{2\tau}e^{-|t|/\tau}$, such that $\G(t)\xrightarrow{\tau\to 0} \delta(t)$, both $\B^\T(-s)$ and $\B(s)$ can be calculated analytically. 
\begin{equation}
\B^\T(-s) = \int_0^{\infty} 2e^{\A^\T t'}\G(t'-s)dt'
=\underbrace{ \frac{1}{\tau} \int_0^s e^{\A^\T t'} e^{-\frac{s-t'}{\tau}}dt' }_{ ({\rm I}) }
+\underbrace{ \frac{1}{\tau} \int_s^\infty e^{\A^\T t'} e^{-\frac{t'-s}{\tau}}dt' }_{ ({\rm II}) }
\end{equation}
We calculate terms $({\rm I})$ and $({\rm II})$ separately:
\begin{align}
({\rm I}) &=  \frac{1}{\tau} \int_0^s e^{\A^\T(s-t')} e^{-\frac{t'}{\tau}} dt'
= \frac{1}{\tau} e^{\A^\T s} \int_0^s e^{ -(\A^\T + \frac{\I}{\tau}) t' } dt'
= \frac{1}{\tau} e^{\A^\T s} \left[  -\left( \A^\T + \frac{\I}{\tau} \right)^{-1}  
e^{ -\left( \A^\T + \frac{\I}{\tau} \right) t' } \right]_0^s \\
&= \frac{1}{\tau} e^{\A^\T s} \left( \A^\T + \frac{\I}{\tau} \right)^{-1}
\left(  \I - e^{ -\left( \A^\T + \frac{\I}{\tau} \right) s } \right)
=e^{\A^\T s} \frac{s}{\tau} \frac{\I - e^{ -\left( \I + \tau \A^\T \right)\frac{s}{\tau} }}{\left( \I + \tau \A^\T \right)\frac{s}{\tau}}
\end{align}
Since the matrix $( \I + \tau \A^\T)$ may in general be non-invertible the result above should be interpreted in terms of evaluating the analytic function $f(x)=\frac{1-e^{-x}}{x}$ for a matrix argument $( \I + \tau \A^\T)\frac{s}{\tau}$.
\begin{align}
({\rm II}) =  \frac{1}{\tau} \int_0^\infty e^{\A^\T (t'+s)} e^{-\frac{t'}{\tau}} dt' 
= \frac{1}{\tau} e^{\A^\T s} \int_0^{\infty} e^{\left( \A^\T -\frac{\I}{\tau} \right) t'} dt'
= -\frac{1}{\tau} e^{\A^\T s} \left( \A^\T -\frac{\I}{\tau} \right)^{-1}
= e^{\A^\T s} ( \I -\tau\A^\T )^{-1}
\end{align}
Similarly, we calculate $\B(s)$:
\begin{equation}
\B(s) = \int_0^{\infty} 2e^{\A t'}\G(t'+s) dt' 
= \frac{1}{\tau} e^{\A t'} e^{-\frac{|t'+s|}{\tau}}
= e^{-\frac{s}{\tau}}\frac{1}{\tau} \int_0^\infty e^{ \left( \A - \frac{\I}{\tau} \right) t' } dt'
=e^{-\frac{s}{\tau}} (\I -\tau \A)^{-1}
\end{equation}
Altogether, the equation for the auto-covariance function reads
\begin{equation}
\A\C(s) + \C(s)\A^\T =
-\D  e^{\A^\T s} \left[ (\I-\tau\A^\T)^{-1} + 
\frac{ \I - e^{-(\I+\tau \A^\T)\frac{s}{\tau}} }{\I+\tau\A^\T} \right]
- e^{-\frac{s}{\tau}} (\I -\tau\A)^{-1} \D
\end{equation}
and in the case $s=0$ the equation reduces to
\begin{equation}
\A\C + \C\A^\T = -[\D (\I-\tau\A^\T)^{-1}  + (\I-\tau\A)^{-1} \D].
\end{equation}

\section{Derivation of the formula for $\AER$}
\label{appendix2}
The area enclosing rates are defined as
\begin{equation}
\AER = \frac{1}{2} \langle \dot{\x} \x^\T - \x \dot{\x}^\T \rangle .
\end{equation}
This antisymmetrized form is used only to make the expression independent of the integration convention in the case of white noise. For colored noise, where no ambiguity in terms of integration occurs, it is enough to calculate $\langle \dot{\x} \x^\T\rangle$. We will confirm this by showing that $\langle \dot{\x} \x^\T\rangle$ itself is antisymmetric. Taking the expression for $\dot{\x}$ from Eq.~\eqref{Eq:Langevin} we get
\begin{equation}
\langle \dot{\x} \x^\T\rangle =
\A \underbrace{\langle \x \x^\T \rangle}_{\C}  
+ \sqrt{ 2\D } \langle \etab(t)\x^\T(t) \rangle.
\end{equation}
Substituting the formal solution for $\x(t)$ we evaluate the second term:
\begin{align}
\sqrt{ 2\D } \langle \etab(t)\x^\T(t) \rangle 
= \int_{-\infty}^t \sqrt{2\D} \langle \etab(t)\etab^\T(t') \rangle \sqrt{2\D} e^{\A^\T(t-t')}dt'
= \int_{-\infty}^t 2\D e^{\A^\T(t-t')} \G(t-t') dt' = \D\B^\T
\end{align}
Combining it with the first term we get
\begin{equation}
\AER = \A\C + \D\B^\T = \frac{1}{2} (\A\C - \C\A^\T) + \frac{1}{2}(\D\B^\T - \B\D).
\label{Eq:supplAERpartial}
\end{equation}
As mentioned before, this expression is already antisymmetric. The first term is identical to the white noise expression, with $\Cw$ replaced by $\C$. The correction term $(\D\B^\T - \B\D)$ contributes to $\AER_{ij}$ only if the degrees of freedom $x_i$, $x_j$ are coupled (directly or indirectly) and if the active noise acts directly on at least one of them. The above expression can be rewritten in an alternative form after making use of $2\C = \B\Cw + \Cw\B^\T$ (Eq.~\eqref{Eq:suppC}) and $2\D = \A\Cw + \Cw\A^\T$ (Eq.~\eqref{Eq:Lyap3}), and the fact that matrices $\A$ and $\B$ commute:
\begin{align}
\AER &= \frac{1}{4} \left[ \A\B\Cw + \A\Cw\B^\T - \B\Cw\A^\T - \Cw\B^\T\A^\T
-\A\Cw\B^\T - \Cw\A^T\B^\T + \B\A\Cw + \B\Cw\A^\T
\right] \\
&=\frac{1}{2} \left[ \B\A\Cw - \Cw\A^\T\B^\T \right] .
\end{align}

\section{Mean velocity field with OU noise}
\label{appendix3}
Consider a linear system driven by OU noise, described by the Langevin equation
\begin{equation}
\begin{pmatrix}
\dot{\x} \\
\dot{\etab}
\end{pmatrix}
=
\begin{pmatrix}
\A & \sqrt{2\D} \\
0 & -\I / \tau
\end{pmatrix}
\begin{pmatrix}
\dot{\x} \\
\dot{\etab}
\end{pmatrix}
+\frac{1}{\tau}
\begin{pmatrix}
0 \\
\xib
\end{pmatrix},
\end{equation}
where $\xib(t)$ is Gaussian white noise satisfying $\langle \xib(t)\xib^\T(t') \rangle = \I\delta(t-t')$. The instantaneous velocity can be then written as $\V=\A\x+\sqrt{2\D}\etab$. The covariance matrix for the whole system $\{ \x, \etab \}$ including the noise dynamics satisfies the Lyapunov equation:
\begin{equation}
\begin{pmatrix}
\A & \sqrt{2\D} \\
0 & -\I / \tau
\end{pmatrix}
\begin{pmatrix}
\C_{\x\x} & \C_{\x\etab} \\
\C_{\etab\x} & \C_{\etab\etab}
\end{pmatrix}
+
\begin{pmatrix}
\C_{\x\x} & \C_{\x\etab} \\
\C_{\etab\x} & \C_{\etab\etab}
\end{pmatrix}
\begin{pmatrix}
\A^\T & 0 \\
\sqrt{2\D} & -\I / \tau
\end{pmatrix}
=
\begin{pmatrix}
0 & 0 \\
0 & -\I / \tau^2
\end{pmatrix}.
\end{equation}
This equation can be splited into three equations for $\{ \C_{\x\x}, \C_{\etab\x}, \C_{\etab\etab} \}$. One of them is
\begin{equation}
-\frac{1}{\tau} \C_{\etab\etab}-\frac{1}{\tau} \C_{\etab\etab} = -\frac{\I}{\tau^2} \Longrightarrow \C_{\etab\etab} = \frac{\I}{2\tau}.
\end{equation}
Another equation, allowing us to find $\C_{\etab\x}$ is
\begin{align}
-\frac{1}{\tau} \C_{\etab\x} + \C_{\etab\x}\A^\T +\C_{\etab\etab}\sqrt{2\D} = 0.
\end{align}
Substituting the derived expression for $\C_{\etab\etab}$ and rearranging the terms we find
\begin{equation}
\C_{\etab\x} = \sqrt{\frac{\D}{2}}(\I - \tau\A^\T)^{-1} = \sqrt{\frac{\D}{2}}\B^\T.
\end{equation}

Because the joint probability distribution $p(\x, \etab)$ has to be multivariate normal, the mean value of the OU noise $\etab$ conditioned on the position $\x$ can be calculated as
\begin{equation}
\langle \etab|\x \rangle =\underbrace{\langle \etab \rangle}_{0} + \C_{\etab\x}(\C_{\x\x})^{-1}\x.
\end{equation}
Consistently with the notation of the main text we will write $\C_{\x\x}\coloneqq\C$. The mean velocity $\V(\x)$ is then obtained as
\begin{align}
\V(\x) &= \langle \V | \x\rangle = \A\x + \sqrt{2\D} \langle \etab|\x\rangle
= (\A + \sqrt{2\D}\C_{\etab\x}(\C_{\x\x})^{-1} )\x 
= \underbrace{(\A + \D\B^T\C^{-1})}_{\Om_{\rm OU}}\x.
\end{align}
Comparing with Eq.~\eqref{Eq:supplAERpartial}, we conclude that with the Ornstein-Uhlenbeck noise the mean velocity can be written as $\langle \V | \x\rangle=\Om_{\rm OU} \x$, where $\Om_{\rm OU}\x = \AER\C^{-1}$, just as for a white noise driven system.
\\

An alternative expression for $\Om_{\rm OU}$ in case of the OU noise can be found by first calculating the exact probability distribution $p(\x,\V)$ and expressing it in terms of the white noise covariance matrix $\Cw$. To this end we followed a perturbative scheme based on a expansion in powers of $\tau^\frac{1}{2}$ that was presented in \cite{Fodor2016a}, employing the identity $\A\Cw+\Cw\A^\T = -2\D$. With linear forces, as in our case, the expansion terminates at order $n=4$ leading to an exact expression:
\begin{align}
p(\x,\V) &= \mathcal{N} \exp \left\{
-\frac{1}{2}
\begin{pmatrix}
\x^\T & \V^\T
\end{pmatrix}
\begin{pmatrix}
\Cw^{-1} + \tau\A^\T\D^{-1}\A & -\tau(\Cw^{-1} + \A^\T\D^{-1}) \\
-\tau(\Cw^{-1} + \D^{-1}\A) & \tau\D^{-1} + \tau^2\Cw^{-1}
\end{pmatrix}
\begin{pmatrix}
\x \\
\V
\end{pmatrix}
\right\} \\
&\coloneqq
\mathcal{N} \exp \left\{
-\frac{1}{2}
\begin{pmatrix}
\x^\T & \V^\T
\end{pmatrix}
\begin{pmatrix}
\C_{\x\x} & \C_{\x\V} \\
\C_{\V\x} & \C_{\V\V}
\end{pmatrix}^{-1}
\begin{pmatrix}
\x \\
\V
\end{pmatrix}
\right\}
,
\end{align}
where $\mathcal{N}$ is a normalization constant.  Additionally, the decomposition of the full covariance matrix using Schur complement gives:
\begin{equation}
[(\C^{-1})_{\V\V}]^{-1} (\C^{-1})_{\V\x} = -\C_{\V\x}(\C_{\x\x})^{-1}.
\end{equation}
Equipped with these results, we can calculate the conditioned mean velocity as:
\begin{align}
\langle \V|\x \rangle &= \C_{\V\x}(\C_{\x\x})^{-1}\x 
= -[(\C^{-1})_{\V\V}]^{-1} (\C^{-1})_{\V\x} \x 
= (\tau\D^{-1} + \tau^2 \Cw^{-1})^{-1} \tau (\Cw^{-1} + \D^{-1}\A) \x \\
&=
(\I + \tau\D\Cw^{-1})^{-1} (\A + \D\Cw^{-1})\x = (\I + \tau\D\Cw^{-1})^{-1}  \Omw\x.
\end{align}
From this we conclude that for a linear system driven by OU noise
\begin{equation}
\langle\V(\x)\rangle = \Om_{\rm OU} \x, \quad \mathrm{with}\quad \Om_{\rm OU} = (\I+\tau\D\Cw^{-1})^{-1}\Omw.
\end{equation}

\end{appendices}
\end{document}